\documentclass[lettersize,journal]{IEEEtran}

\ifCLASSINFOpdf
  \usepackage[pdftex]{graphicx}
  \DeclareGraphicsExtensions{.pdf,.jpeg,.png}
\else
  \usepackage[dvips]{graphicx}
\fi
\ifCLASSOPTIONcompsoc
\usepackage[caption=false, font=normalsize, labelfont=sf, textfont=sf]{caption,subfig}
\else
\usepackage[caption=false, font=footnotesize]{subfig}
\fi
\usepackage{epstopdf}
\usepackage[cmex10]{amsmath}
\usepackage{amssymb}
\usepackage{wasysym}
\usepackage{array}
\usepackage{cite}
\usepackage{color}
\usepackage{url}
\usepackage{bm}
\usepackage{enumerate}
\usepackage{epsfig,latexsym}%amsmath
\usepackage{flushend}
\usepackage{cite}
\usepackage{hyperref}
\usepackage{algorithm}
\usepackage{algorithmic}
\usepackage{diagbox} % ¼ÓÔغê°ü
\usepackage{booktabs}
\usepackage{stfloats}
\usepackage{cases}

\makeatletter

\renewcommand{\maketag@@@}[1]{\hbox{\m@th\normalsize\normalfont#1}}%

\makeatother

\begin{document}
\title{Hybrid Beamforming Design for Covert mmWave MIMO with Finite-Resolution DACs}

\author{Wei~Ci,~\IEEEmembership{Graduate Student Member,~IEEE}, Chenhao~Qi,~\IEEEmembership{Senior Member,~IEEE} and Xiaohu~You,~\IEEEmembership{Fellow,~IEEE}
        % <-this % stops a space		
\thanks{Part of this work has been accepted for presentation at the 2024 lEEE Global Communications Conference (GLOBECOM), Cape Town, South Africa, Dec. 2024~\cite{Ci2024Finite}. (\textit{Corresponding author: Chenhao Qi})}
\thanks{Wei~Ci, Chenhao~Qi and Xiaohu~You are with the National Mobile Communications Research Laboratory, School of Information Science and Engineering, Southeast University, Nanjing 210096, China (e-mail: \{cw99,qch,xhyu\}@seu.edu.cn).}}

\markboth{Accepted By IEEE Journal on Selected Areas in Communications}{}

\maketitle



\begin{abstract}
We investigate hybrid beamforming design for covert millimeter wave multiple-input multiple-output systems with finite-resolution digital-to-analog converters (DACs), which impose practical hardware constraints not yet considered by the existing works and have negative impact on the covertness. Based on the additive quantization noise model, we derive the detection error probability of the warden considering finite-resolution DACs. Aiming at maximizing the sum covert rate (SCR) between the transmitter and legitimate users, we design hybrid beamformers subject to power and covertness constraints. To solve this nonconvex joint optimization problem, we propose an alternating optimization (AO) scheme based on fractional programming, quadratic transformation, and inner majorization-minimization methods to iteratively optimize the analog and digital beamformers. To reduce the computational complexity of the AO scheme, we propose a vector-space based heuristic (VSH) scheme to design the hybrid beamformer. We prove that as the number of antennas grows to be infinity, the SCR in the VSH scheme can approach the channel mutual information. Simulation results show that the AO and VSH schemes outperform the existing schemes and the VSH scheme can be used to obtain an initialization for the AO scheme to speed up its convergence.
\end{abstract}

\begin{IEEEkeywords}
	Covert communications, digital-to-analog converter (DAC), hybrid beamforming, millimeter wave (mmWave) communications, multiuser communications.
\end{IEEEkeywords}

\section{Introduction}
Wireless communications pose significant challenges in terms of security and privacy due to their broadcast nature. Traditional secure communications primarily focus on preventing messages from being decoded by potential eavesdroppers. However, in specific scenarios such as battlefield environments, even the regular wireless communications can expose us to the enemy and result in fatal danger. This leads to the emergence of covert communication techniques to achieve enhanced security. Nevertheless, achieving covert communications in an open wireless environment poses significant challenges. Fortunately, millimeter wave (mmWave) communications inherently provide the covertness due to the sparse scattering characteristics~\cite{Hu2021DRL,Qi2020HBT}. By employing massive antenna arrays, highly directional beams can be formed to effectively mitigate energy leakage and decrease the risk of interception by unauthorized parties. Consequently, balancing communication performance and covertness becomes a primary concern in mmWave multiple-input multiple-output (MIMO) systems. In a typical wireless covert communication system, a legitimate transmitter aims to communicate with a legitimate user without being detected by a warden. However, for the additive white Gaussian noise (AWGN) channel, the square root law is proved in~\cite{Bash2013Limits} that only $\mathcal{O}(\sqrt{n})$ bits can be transmitted in $n$ channel uses, indicating that the covert bits per channel use decreases to zero if $n$ grows to be infinity. Similar findings are observed in discrete memoryless channels~\cite{Bloch2016Covert}, broadcast channels~\cite{AKS2019Embedding} and multiple access channels~\cite{AKS2019Covert}. Therefore, from the warden's perspective, various uncertainties in wireless environments, such as noise~\cite{Goeckel2016covert} and channel fading~\cite{Shahzad2017Fading}, should be considered to achieve a positive covert rate. Furthermore, artificial jamming signals are introduced to disrupt the warden's detection so that the covertness can be further enhanced~\cite{Sobers2017Uninformed,Shahzad2018Achieving,Zheng2021Wireless}.

All the works mentioned above only consider the single-antenna transmitter regime. Different from introducing uncertainties for the covertness, we can also mitigate the energy leakage to the warden through elaborately designed beamformers to enhance the covertness performance~\cite{Ma2021Robust,Jamali2022Covert,Zhang2021Beamtraining,Xing2023covert}. Specifically, for two cases that the transmitter knows perfect or imperfect channel state information (CSI) of the warden, the zero-forcing and robust beamformers are proposed to maximize the covert rate~\cite{Ma2021Robust}. In~\cite{Jamali2022Covert}, beamforming for covert communications in mmWave bands is considered for a transmitter with two independent arrays generates two beams, where one beam towards the legitimate user is for data transmission and the other beam towards the warden carries a jamming signal. Since the energy leakage in the beam training phase may bring potential risks, the covertness in both beam training and data transmission stages is investigated by jointly optimizing transmit power and beam training durations to maximize the covert throughput~\cite{Zhang2021Beamtraining}. Since the ideal beam pattern used in~\cite{Zhang2021Beamtraining} is impractical, a discrete Fourier transform codebook is used for beam training in the covert communication system~\cite{Xing2023covert}. Additionally, the advancement of massive antenna arrays leads to the emergence of hybrid beamforming techniques~\cite{Cai2020Secure,Chen2020twostep}, which can also be used in covert communications. A joint hybrid analog, digital beamforming and jamming design algorithm is proposed for a covert communication system with a full-duplex receiver, where the detection error probabilities of the warden are derived for both single and multiple data stream cases~\cite{Wang2022covertRate}.

However, the aforementioned works mainly investigate the ideal beamforming design with infinite-resolution digital-to-analog converters (DACs) or analog-to-digital converters (ADCs) without considering quantization noise. On one hand, employing high-resolution DACs or ADCs results in significant power consumption~\cite{Orhan2015LowPower}. On the other hand, using low-resolution DACs or ADCs introduces significant quantization noise, which has negative impact on both the communication and covertness performance. Therefore, it is crucial to investigate beamforming techniques with regard of the resolution of the DACs or ADCs~\cite{Abbas2017Lowresolution,Xu2019Secure,Wu2023secure}. Specifically, for a receiver with finite-resolution ADCs, three types of beamforming design schemes corresponding to analog, hybrid, and fully-digital architectures are proposed to maximize the achievable rate, where the energy efficiency performance of the proposed schemes is also analyzed~\cite{Abbas2017Lowresolution}. Furthermore, beamforming design schemes with finite-resolution DACs for secure communications are investigated, where the analog beamformer, digital beamformer and artificial jammer are jointly optimized to maximize the secrecy rate~\cite{Xu2019Secure,Wu2023secure}. Since the quantization noise introduced by finite-resolution DACs also has a negative impact on the covert communications, we design hybrid beamformers with finite-resolution DACs in a covert communication system. The main contributions of this paper are summarized as follows, where the first and second points are included in the conference paper~\cite{Ci2024Finite}.
\begin{itemize}
	\item[1)] We investigate hybrid beamforming design for covert mmWave MIMO systems with finite-resolution DACs, which are practical hardware constraints not yet considered by the
	existing works. Based on the additive quantization noise (AQN) model, we derive the detection error probability of the warden to evaluate the level of the covertness in this system. Aiming to maximize the sum covert rate (SCR) between the transmitter and legitimate users, we establish an optimization problem for the hybrid beamforming design subject to power and covertness constraints.
	\item[2)] To solve the non-convex optimization problem, we propose an alternating optimization (AO) scheme. Initially, we employ quadratic and Lagrangian dual transformations to decompose the optimization problem into four subproblems, which are iteratively solved until convergence. For the subproblem of designing the analog beamformer, we first transform it into a quadratically constrained quadratic programming (QCQP) problem with the extra constant-modulus constraints. Then, we solve the transformed problem using an inner majorization-minimization (iMM) method. Additionally, we transform the subproblem of designing the digital beamformer into a standard QCQP convex problem, which is solved by the interior-point method.
	\item[3)] To reduce the computational complexity of the AO scheme, we propose a vector-space based heuristic (VSH) scheme for hybrid beamforming design. We prove that the SCR in the VSH scheme can approach the channel mutual information as the number of antennas grows to be infinity. Moreover, simulation results demonstrate that the VSH scheme can be used to obtain an initialization for the AO scheme to speed up its convergence.
\end{itemize}

The rest of this paper is structured as follows. The system model and problem formulation are presented in Section~\ref{section2}. In Section~\ref{section3}, we propose the hybrid beamforming design based on the AO scheme. The VSH scheme is proposed for hybrid beamforming design in Section~\ref{section4}. Simulation results are presented in Section~\ref{section5}. Finally, this paper is concluded in Section~\ref{section6}.

The notations are defined as follows. Symbols for vectors (lower case) and matrices (upper case) are in boldface. $(\cdot)^*$, $(\cdot)^{\rm T}$ and $(\cdot)^{\rm H}$ denote the conjugate, transpose and conjugate transpose, respectively. $[\boldsymbol{a}]_n$, $\|\boldsymbol{a}\|_{\rm 2}$, $[\boldsymbol{A}]_{m,n}$, $[\boldsymbol{A}] _{:,m}$, $[\boldsymbol{A}]_{:,n:m}$ and $\|\boldsymbol{A}\|_{\rm F}$ denote the $n$th entry, the $\ell_2$ norm of the vector $\boldsymbol{a}$, the entry on the $m$th row and $n$th column, the $m$th column, the columns from $n$th to $m$th and the Frobenius norm of the matrix $\boldsymbol{A}$, respectively. $\boldsymbol{A}\succeq 0$ indicates that  $\boldsymbol{A}$ is a positive semi-definite matrix. $\boldsymbol{I}_{L}$ denotes an $L \times L$ identity matrix. The functions ${\rm vec}(\cdot)$, ${\rm vec}^{-1}(\cdot)$, ${\rm Tr}(\cdot)$ and ${\rm diag}(\cdot)$ denote the vectorization, the inverse operation of vectorization, trace and diagonal elements of a matrix, respectively. ${\rm Re}(\cdot)$ and $\angle(\cdot)$ denote the real part and the angle of a complex number, respectively. $\mathbb{E}(\cdot)$ denotes the statistical expectation. ${\rm Diag}(a_1,a_2,\cdots,a_N)$ denotes the diagonal matrix whose diagonal elements are $a_1,a_2,\cdots,a_N$. ${\rm min}(a_1,a_2,\cdots,a_N)$ denotes the minimum value among $a_1,a_2,\cdots,a_N$. $\mathcal{CN}(\boldsymbol{\mu},\boldsymbol{\Sigma})$ denotes a complex Gaussian distribution with mean $\boldsymbol{\mu}$ and covariance matrix $\boldsymbol{\Sigma}$. Symbols $j$, $\mathbb{R}$, $\mathbb{C}$ and $\otimes$ denote the square root of $-1$, the set of real-valued numbers, complex-valued numbers and the Kronecker product, respectively.

\section{System Model and Problem Formulation}\label{section2}
\subsection{System Model}
\begin{figure}[!t]
	\centering
	\includegraphics[width=90mm]{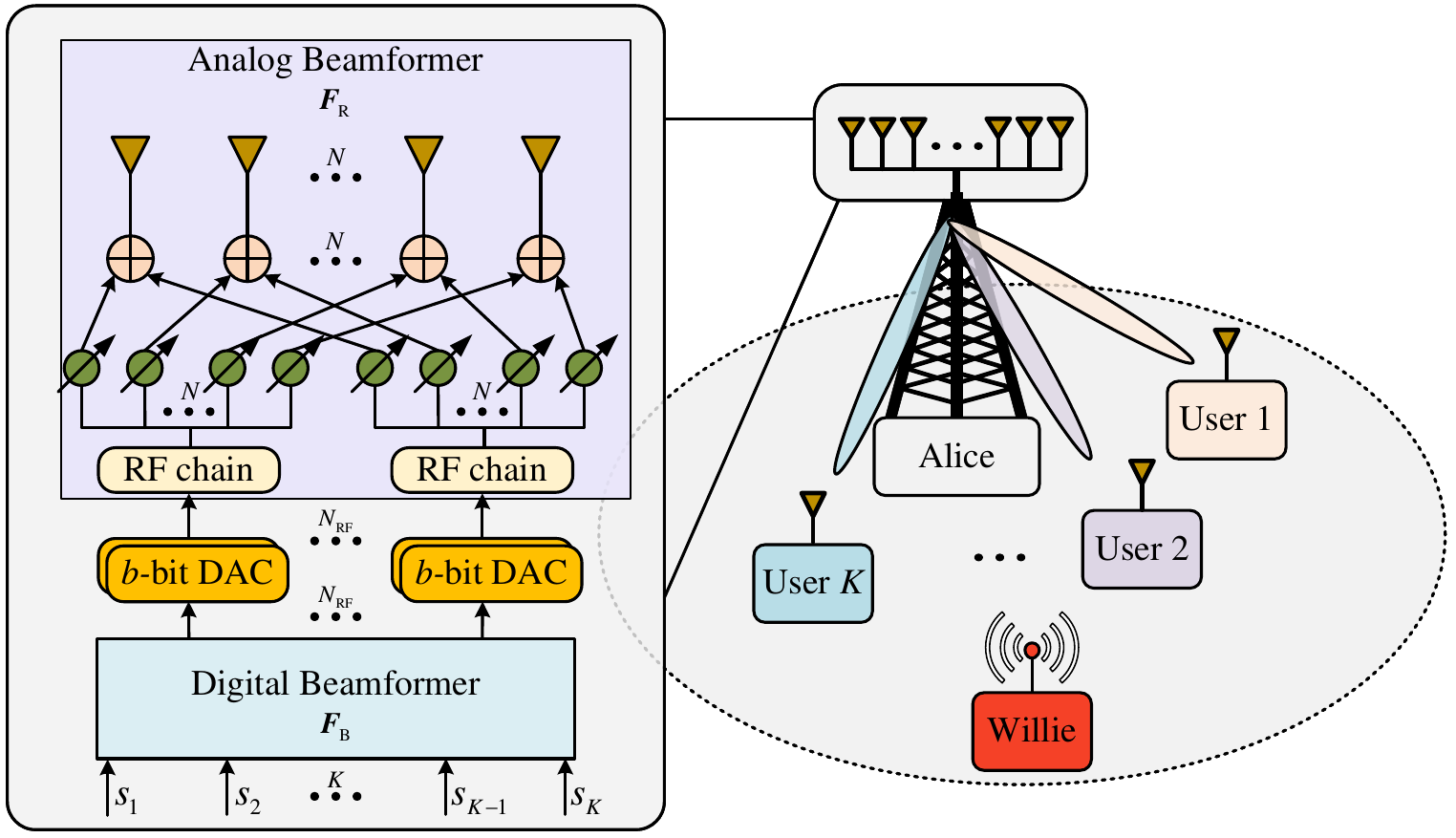} 
	\caption{Illustration of the system model.}
	\label{fig:systemmodel}
	\vspace{-0.4cm}
\end{figure}

We consider a covert mmWave MIMO communication system as shown in Fig.~\ref{fig:systemmodel}. The base station named Alice simultaneously communicates with $K$ legitimate users, meanwhile a warden named Willie attempts to detect the existence of the communications. Alice uses a fully-connected hybrid beamforming architecture with $N$ antennas and $N_{\rm RF}$ radio frequency (RF) chains, where the antennas are placed in uniform linear arrays with half wavelength intervals and each RF chain is connected to a finite-resolution DAC. To reduce the hardware complexity and ensure the performance of massive antenna array communications, we set $N_{\rm RF}=K \ll N$. The $K$ legitimate users and Willie are all equipped with a single antenna for simplicity. 

The transmitted information symbols of $K$ data streams from Alice are firstly precoded by a baseband digital beamformer $\boldsymbol{F}_{\rm B}\triangleq [\boldsymbol{f}_{{\rm B},1},\cdots,\boldsymbol{f}_{{\rm B},K}]\in \mathbb{C}^{K\times K}$, then converted by finite-resolution DACs. The baseband transmitted signal can be expressed as
\begin{equation}\label{Basebandsignal}
	\boldsymbol{x}_{\rm b} = \mathbb{Q}\big(\sum_{k=1}^{K}\boldsymbol{f}_{{\rm B},k}s_k\big),
\end{equation}
where $s_k\in \mathbb{C}$ for $k = 1,2,\cdots,K$ is the information symbol transmitted to the $k$th user. $\mathbb{Q}(\cdot)$ denotes the quantization function imposed by the finite-resolution DACs. Additionally, we denote $\boldsymbol{s}\triangleq [s_1,s_2,\cdots,s_k]^{\rm T}\sim \mathcal{CN}(\boldsymbol{0},\boldsymbol{I}_K)$, indicating that the symbols from different data streams are independent of each other. To derive a tractable expression for \eqref{Basebandsignal}, we use the AQN model~\cite{Roth2017lowbeamforming,Fletcher2007AQNM} to approximate the output with an linear form as
\begin{equation}
	\boldsymbol{x}_{\rm b} \approx (1-\beta) \sum_{k=1}^{K}\boldsymbol{f}_{{\rm B},k}s_k + \boldsymbol{\eta}_{\rm q},
\end{equation}
where $\beta$ is the quantization distortion parameter and it depends on the resolution of the DACs. Let $b$ denote the number of quantized bits of the DACs. Specifically, if we set $b=1,2,\cdots,5$, the values of $\beta$ are 0.3634, 0.1175, 0.03454, 0.009497 and 0.002499, respectively. When $b>5$, $\beta$ can be approximated by $\frac{\pi\sqrt{3}}{2}2^{-2b}$. As $b$ grows to be infinity, $\beta$ will be 0 and it means that there is no quantization distortion. $\boldsymbol{\eta}_{\rm q} \sim \mathcal{CN}(\boldsymbol{0},\boldsymbol{R}_{\rm q})$ denotes the quantization noise and its covariance matrix can be expressed as
\begin{equation}\label{Rq}
	\boldsymbol{R}_{\rm q} = \beta(1-\beta){\rm diag}\Big(\sum_{k=1}^{K}\boldsymbol{f}_{{\rm B},k}\boldsymbol{f}_{{\rm B},k}^{\rm H}\Big).
\end{equation}
After being up-converted to the RF domain, the transmitted signals are precoded by an analog beamformer $\boldsymbol{F}_{\rm R}\in \mathbb{C}^{N\times K}$. Therefore, the received signal by the $k$th legitimate user can be expressed as
\begin{equation}\label{ykwithQ}
	y_k = \boldsymbol{h}_k^{\rm H}\boldsymbol{F}_{\rm R}\big(\sum_{l=1}^{K}(1-\beta)\boldsymbol{f}_{{\rm B},l}s_l+\boldsymbol{\eta}_{\rm q}\big)+\overline{\eta}_k,
\end{equation}
where $\overline{\eta}_k\sim \mathcal{CN}(0,\sigma_k^2)$ denotes the AWGN at the $k$th legitimate user. $\boldsymbol{h}_k$ denotes the channel vector between Alice and the $k$th user and can be expressed as
\begin{equation}\label{channelhk}
	\boldsymbol{h}_k = \sqrt{\frac{N}{D_k}}\Big(\alpha_k^{(0)}\boldsymbol{a}(N,\theta_k^{(0)}) + \sum_{d=1}^{D_k-1}\alpha_k^{(d)}\boldsymbol{a}(N,\theta_k^{(d)})\Big),
\end{equation}
where $D_k$ denotes the total number of channel paths between Alice and $k$th user. $\alpha_k^{(0)}$ and $\alpha_k^{(d)}$ for $d = 1,2,\cdots,D_k-1$ denote the channel gain for the line-of-sight (LoS) and the $d$th non-line-of-sight (NLoS) channel paths, respectively. $\theta_k^{(0)}$ and $\theta_k^{(d)}$ for $d = 1,2,\cdots,D_k-1$ denote the channel angle-of-departure (AoD) for the LoS and the $d$th NLoS channel paths, respectively. Moreover, $\boldsymbol{a}(N,\theta)$ is the normalized array response and can be expressed as
\begin{equation}\label{arraryresponse}
	\boldsymbol{a}(N,\theta) = \frac{1}{\sqrt{N}}\big[1,e^{j\pi\sin(\theta)},\cdots,e^{j\pi(N-1)\sin(\theta)}\big]^{\rm T}.
\end{equation}
Similarly, the channel vector between Alice and Willie can be expressed as
\begin{equation}
	\boldsymbol{h}_{\rm w} = \sqrt{\frac{N}{D_{\rm w}}}\Big(\alpha_{\rm w}^{(0)}\boldsymbol{a}(N,\theta_{\rm w}^{(0)}) + \sum_{d=1}^{D_{\rm w}-1}\alpha_{\rm w}^{(d)}\boldsymbol{a}(N,\theta_{\rm w}^{(d)})\Big),
\end{equation}
where $D_{\rm w},\alpha_{\rm w}$ and $\theta_{\rm w}$ are distinguished with $D_k,\alpha_k$ and $\theta_k$ in \eqref{channelhk}. Note that $\boldsymbol{h}_{\rm w} \sim \mathcal{CN}(0,\boldsymbol{\Omega}_{\rm w})$ and the covariance matrix $\boldsymbol{\Omega}_{\rm w}$ can be expressed as
\begin{equation}
	\boldsymbol{\Omega}_{\rm w} \triangleq \frac{N}{D_{\rm w}}\sum_{d=0}^{D_{\rm w}-1} \mathbb{E}|\alpha_{\rm w}^{(d)}|^2\boldsymbol{a}(N,\theta_{\rm w}^{(d)})\boldsymbol{a}(N,\theta_{\rm w}^{(d)})^{\rm H}.
\end{equation}

In this paper, we assume that Alice knows the instantaneous CSI of $\boldsymbol{h}_k$ for $k = 1,2,\cdots,K$, which can be obtained via uplink training for channel estimation. However, Willie is usually a passive node and obtaining the instantaneous CSI for $\boldsymbol{h}_{\rm w}$ is almost impossible. Therefore, we assume that only the statistical CSI $\boldsymbol{\Omega}_{\rm w}$ is available to Alice~\cite{Wang2022covertRate}. For the worst case, we assume that Willie knows all the instantaneous CSIs of $\boldsymbol{h}_{\rm w}$ and $\boldsymbol{h}_k$ for $k = 1,2,\cdots,K$.

\subsection{Detection Performance of Willie}
The signal detection process of Willie can be formulated as a binary hypothesis testing problem. Specifically, the hypothesis $\mathcal{H}_0$ represents that Alice remains silent, while the hypothesis $\mathcal{H}_1$ represents that Alice is communicating with the $K$ legitimate users. Then the received signal at Willie in the $t$th time slot can be derived as
\begin{align}
	\mathcal{H}_0:y_{\rm w}[t]&=\overline{\eta}_{\rm w}[t], \label{H0}\\
	\mathcal{H}_1:y_{\rm w}[t]&=\boldsymbol{h}_{\rm w}^{\rm H}\boldsymbol{F}_{\rm R}\big(\sum_{l=1}^{K}(1-\beta)\boldsymbol{f}_{{\rm B},l}s_l[t]+\boldsymbol{\eta}_{\rm q}[t]\big)+\overline{\eta}_{\rm w}[t], \label{H1}
\end{align}
where $\overline{\eta}_{\rm w}[t]\sim \mathcal{CN}(0,\sigma_{\rm w}^2)$ denotes the AWGN received by Willie in the $t$th time slot. For representation simplicity, we define $\mathcal{D}_0$ and $\mathcal{D}_1$ to represent the decisions of Willie under $\mathcal{H}_0$ and $\mathcal{H}_1$, respectively. Moreover, $\mathcal{P}_{\rm FA}\triangleq {\rm Pr}(\mathcal{D}_1|\mathcal{H}_0)$ denotes the false alarm probability and represents that Willie performs inference $\mathcal{D}_1$ but $\mathcal{H}_0$ holds. $\mathcal{P}_{\rm MD}\triangleq {\rm Pr}(\mathcal{D}_0|\mathcal{H}_1)$ denotes the missed detection probability and represents that Willie performs inference $\mathcal{D}_0$ but $\mathcal{H}_1$ holds. Similar to \cite{Shahzad2017Fading,Wang2022covertRate}, since Willie is unaware of when Alice will transmit, we assume that Alice transmits with equal transmit prior probabilities, i.e., ${\rm Pr}(\mathcal{H}_0) = {\rm Pr}(\mathcal{H}_1) = \frac{1}{2}$. Therefore, we can derive the detection error probability of Willie as
\begin{equation}
	\mathcal{P}_{\rm e} = \mathcal{P}_{\rm FA} + \mathcal{P}_{\rm MD},
\end{equation}
$\mathcal{P}_{\rm e} = 1$ means that Willie always makes error detection and communication covertness is achieved perfectly in this case. The detailed derivation of $\mathcal{P}_{\rm e}$ will be included in Section~\ref{section3}.

\subsection{Problem Formulation}
With the derivation of \eqref{Rq} and \eqref{ykwithQ}, we can define ${\rm SIQNR}_k$ as the signal to interference, quantization distortion and noise ratio for the $k$th user, which can be expressed as \eqref{SIQNR} at the top of this page. Then the SCR can be given by
\setcounter{equation}{12}
\begin{equation}\label{Rsum}
	R_{\rm sum} = \sum_{k=1}^K \log(1+{\rm SIQNR}_k).
\end{equation}

\setcounter{equation}{11}
\begin{figure*}[ht]
	\vspace*{5pt}
	\begin{equation}\label{SIQNR}
		{\rm SIQNR}_k = \frac{(1-\beta)^2|\boldsymbol{h}_k^{\rm H}\boldsymbol{F}_{\rm R}\boldsymbol{f}_{{\rm B},k}|^2}{(1-\beta)^2\sum_{l=1,l\neq k}^{K}|\boldsymbol{h}_k^{\rm H}\boldsymbol{F}_{\rm R}\boldsymbol{f}_{{\rm B},l}|^2+\boldsymbol{h}_k^{\rm H}\boldsymbol{F}_{\rm R}\boldsymbol{R}_{\rm q}\boldsymbol{F}_{\rm R}^{\rm H}\boldsymbol{h}_k+ \sigma_k^2}, k = 1,2,\cdots,K.
	\end{equation}
    \rule[0pt]{18.3cm}{0.06em}
\end{figure*}
\setcounter{equation}{13}

The objective of this paper is to maximize the SCR by jointly optimizing the analog beamformer $\boldsymbol{F}_{\rm R}$ and the digital beamformer $\boldsymbol{F}_{\rm B}$, subject to the following constraints: (1) The constant modulus constraints on the elements of analog beamformer, i.e., $\big|[\boldsymbol{F}_{\rm R}]_{n,m}\big| = 1$ for $n = 1,\cdots, N,$ and $m=1,\cdots,K$; (2) The transmit power constraint under the AQN model, i.e., \begin{footnotesize}$\mathbb{E}\Big(\Big\|\boldsymbol{F}_{\rm R}\big((1-\beta)\boldsymbol{F}_{\rm B}\boldsymbol{s}+\boldsymbol{\eta}_{\rm q}\big)\Big\|_{\rm F}^2\Big) \leq P_{\rm max}$\end{footnotesize}, where $P_{\rm max}$ is the maximum transmit power; (3) The covertness constraint from Alice's perspective, $\mathbb{E}_{\boldsymbol{h}_{\rm w}}\big(\mathcal{P}_{\rm e}\big) \geq 1-\epsilon$, where $\epsilon$ denotes the predetermined level of the covertness. Taking the above three constraints into consideration, we can express the optimization problem as
\begin{subequations}\label{goalformulation}
	\begin{align}
		\underset{\boldsymbol{F}_{\rm R},\boldsymbol{F}_{\rm B}}{\max}\ &R_{\rm sum} \label{goalformulation:sub1}\\
		\mathrm{s.t.\ }~~&\big|[\boldsymbol{F}_{\rm R}]_{n,m}\big| = 1, n = 1,\cdots, N, m=1,\cdots,K, \label{goalformulation:sub2}\\
		~&\mathbb{E}\bigg(\Big\|\boldsymbol{F}_{\rm R}\big((1-\beta)\boldsymbol{F}_{\rm B}\boldsymbol{s}+\boldsymbol{\eta}_{\rm q}\big)\Big\|_{\rm F}^2\bigg) \leq P_{\rm max}, \label{goalformulation:sub3}\\
		~&\mathbb{E}_{\boldsymbol{h}_{\rm w}}(\mathcal{P}_{\rm e}) \geq 1-\epsilon. \label{goalformulation:sub4}
	\end{align}
\end{subequations}

Unfortunately, \eqref{goalformulation} is a non-convex problem and challenging to handle owing to the coupling of $\boldsymbol{F}_{\rm R}$ and $\boldsymbol{F}_{\rm B}$. To efficiently solve this problem, we initially decompose this problem using quadratic transform and Lagrangian dual transform. Subsequently, we will propose the AO and VSH schemes to design the hybrid beamformer in Section \ref{section3} and Section \ref{section4}, respectively.

\section{AO Scheme for Hybrid Beamforming Design}\label{section3}
In this section, we will propose the AO scheme to solve \eqref{goalformulation}. Specifically, the procedures for transforming \eqref{goalformulation} into a solvable form are presented in Section~\ref{3_A}. The detailed AO scheme for hybrid beamforming design is proposed in Section~\ref{3_B}. The complexity analysis is presented in Section~\ref{3_C}.

\subsection{Problem Transformation}\label{3_A}
We first derive the expression of the power constraint \eqref{goalformulation:sub3}. By substituting \eqref{Rq} into \eqref{goalformulation:sub3} and using the fact that $\|\boldsymbol{A}\|_{\rm F}^2 = {\rm Tr}(\boldsymbol{A}\boldsymbol{A}^{\rm H})$, \eqref{goalformulation:sub3} can be transformed as
\begin{equation}\label{powercstr}
	{\rm Tr}\Big(\boldsymbol{F}_{\rm R}\big((1-\beta)^2\sum_{l=1}^{K}\boldsymbol{f}_{{\rm B},l}\boldsymbol{f}_{{\rm B},l}^{\rm H}+\boldsymbol{R}_{\rm q}\big)\boldsymbol{F}_{\rm R}^{\rm H}\Big) \leq P_{\rm max}.
\end{equation}

Then, we derive the expression of $\mathcal{P}_{\rm e}$ in \eqref{goalformulation:sub4}. Let $\mathbb{P}_0 \triangleq f(\boldsymbol{y}_{\rm w}[t]|\mathcal{H}_0)$ and $\mathbb{P}_1 \triangleq f(\boldsymbol{y}_{\rm w}[t]|\mathcal{H}_1)$ denote the probability distribution of the received signal at Willie under $\mathcal{H}_0$ and $\mathcal{H}_1$ in one time slot, respectively. Moreover, we can obtain $f(\boldsymbol{y}_{\rm w}[t]|\mathcal{H}_0) = \mathcal{CN}(\boldsymbol{0}, \sigma_{\rm w}^2)$ and $f(\boldsymbol{y}_{\rm w}[t]|\mathcal{H}_1) = \mathcal{CN}\Big(\boldsymbol{0}, \boldsymbol{h}_{\rm w}^{\rm H}\boldsymbol{F}_{\rm R}\big((1-\beta)^2\sum_{l=1}^{K}\boldsymbol{f}_{{\rm B},l}\boldsymbol{f}_{{\rm B},l}^{\rm H}+\boldsymbol{R}_{\rm q}\big)\boldsymbol{F}_{\rm R}^{\rm H}\boldsymbol{h}_{\rm w}+\sigma_{\rm w}^2\Big)$, respectively. Similar to \cite{Zhao2024SDMA}, we suppose that Willie uses $T$ consecutive time slots for detection and the corresponding joint probability distribution for the $T$ independent observations under $\mathcal{H}_0$ and $\mathcal{H}_1$ can be denoted as
\begin{align}
	&\mathbb{P}^{T}_0 \triangleq \prod_{t=1}^{T} f(\boldsymbol{y}_{\rm w}[t]|\mathcal{H}_0), \label{PT1}\\
	&\mathbb{P}^{T}_1 \triangleq \prod_{t=1}^{T} f(\boldsymbol{y}_{\rm w}[t]|\mathcal{H}_1). \label{PT0}
	\end{align}

Since Willie is aware of $\boldsymbol{h}_{\rm w}$ and we consider the worst case that Willie knows $\boldsymbol{F}_{\rm R}$, $\boldsymbol{F}_{\rm B}$ and $\boldsymbol{R}_{\rm q}$ in advance, Willie can perform the optimal test~\cite{Bash2013Limits} to minimize the detection error probability and the corresponding minimum $\widehat{\mathcal{P}}_{\rm e}$ can be derived as
\begin{equation}\label{Pehat}
	\widehat{\mathcal{P}}_{\rm e} = 1 - \mathcal{V}(\mathbb{P}^{T}_1, \mathbb{P}^{T}_0),
\end{equation}
where $\mathcal{V}(\mathbb{P}^{T}_1, \mathbb{P}^{T}_0)$ denotes the total variation distance between $\mathbb{P}^{T}_1$ and $\mathbb{P}^{T}_0$. Since it is difficult to further analyze the expressions of $\mathcal{V}(\mathbb{P}^{T}_1, \mathbb{P}^{T}_0)$, we can use the Pinsker's inequality~\cite{Bash2013Limits} to obtain a tractable upper bound as
\begin{equation}\label{VT2D}
	\mathcal{V}(\mathbb{P}^{T}_1, \mathbb{P}^{T}_0) \leq \sqrt{\frac{1}{2}\mathcal{D}(\mathbb{P}^{T}_1, \mathbb{P}^{T}_0)},
\end{equation}
where $\mathcal{D}(\mathbb{P}^{T}_1, \mathbb{P}^{T}_0)$ denotes the Kullback-Leibler (KL) divergence from $\mathbb{P}^{T}_1$ to $\mathbb{P}^{T}_0$ and can be derived via the following proposition.

\emph{Proposition 1:} $\mathcal{D}(\mathbb{P}^{T}_1, \mathbb{P}^{T}_0)$ can be expressed as
\begin{equation}
	\mathcal{D}(\mathbb{P}^{T}_1, \mathbb{P}^{T}_0) = T\big(\xi - \ln(1+\xi)\big),
\end{equation}
where 
\begin{equation}\label{xi}
	\xi = \frac{\boldsymbol{h}_{\rm w}^{\rm H}\boldsymbol{F}_{\rm R}\big((1-\beta)^2\sum_{l=1}^{K}\boldsymbol{f}_{{\rm B},l}\boldsymbol{f}_{{\rm B},l}^{\rm H}+\boldsymbol{R}_{\rm q}\big)\boldsymbol{F}_{\rm R}^{\rm H}\boldsymbol{h}_{\rm w}}{\sigma_{\rm w}^2}.
\end{equation}

\emph{Proof:} We first denote $\tau_0^2 \triangleq \sigma_{\rm w}^2$ and $\tau_1^2 \triangleq \boldsymbol{h}_{\rm w}^{\rm H}\boldsymbol{F}_{\rm R}\big((1-\beta)^2\sum_{l=1}^{K}\boldsymbol{f}_{{\rm B},l}\boldsymbol{f}_{{\rm B},l}^{\rm H}+\boldsymbol{R}_{\rm q}\big)\boldsymbol{F}_{\rm R}^{\rm H}\boldsymbol{h}_{\rm w}+\sigma_{\rm w}^2$, then we can obtain $\mathbb{P}_0 \triangleq \frac{1}{\pi \tau_0^2} {\rm exp}(-\frac{|x|^2}{\tau_0^2})$ and $\mathbb{P}_1 \triangleq \frac{1}{\pi \tau_1^2} {\rm exp}(-\frac{|x|^2}{\tau_1^2})$. According to the definition of KL divergence, we can obtain
\begin{align}
	\mathcal{D}(\mathbb{P}_1, \mathbb{P}_0) &= \mathbb{E}_{\mathbb{P}_1}(\ln\mathbb{P}_1 - \ln\mathbb{P}_0) \nonumber\\
	&= \mathbb{E}_{\mathbb{P}_1}\big(\ln\frac{\tau_0^2}{\tau_1^2}+(\frac{1}{\tau_0^2}-\frac{1}{\tau_1^2})|x|^2\big) \nonumber\\
	&= \ln\frac{\tau_0^2}{\tau_1^2} + \frac{\tau_1^2}{\tau_0^2} - 1 \nonumber\\
	&= \xi - \ln(1+\xi).
\end{align}
Using the fact that $\mathcal{D}(\mathbb{P}_1^{T}, \mathbb{P}_0^{T}) = T\mathcal{D}(\mathbb{P}_1, \mathbb{P}_0)$~\cite[Eq. (2.67)]{EIT2002}, Proposition 1 is therefore proved. $\hfill\blacksquare$

From Alice's perspective, considering \eqref{goalformulation:sub4}, \eqref{Pehat} and \eqref{VT2D}, the covertness constraint can be replaced by
\begin{equation}\label{Ehw_convert}
	\mathbb{E}_{\boldsymbol{h}_{\rm w}}\Big(\sqrt{\frac{1}{2}\mathcal{D}(\mathbb{P}^{T}_1, \mathbb{P}^{T}_0)}\Big) \leq \epsilon.
\end{equation}
By using similar techniques to \cite{Wang2022covertRate} and \cite{Ma2022covertrand}, we can derive an upper bound for the left hand of (23) to ensure a safer scenario for the covertness, which can be expressed as
	\begin{align}\label{Ehw}
		&~~~~\mathbb{E}_{\boldsymbol{h}_{\rm w}}\Big(\sqrt{\frac{1}{2}\mathcal{D}(\mathbb{P}^{T}_1, \mathbb{P}^{T}_0)}\Big) \nonumber\\&\overset{(a)}{\leq} \mathbb{E}_{\boldsymbol{h}_{\rm w}}\big(\frac{\sqrt{T}}{2}\xi\big) \nonumber\\
		&= \frac{\sqrt{T}}{2\sigma_{\rm w}^2}\mathbb{E}_{\boldsymbol{h}_{\rm w}}\!\bigg(\!\boldsymbol{h}_{\rm w}^{\rm H}\Big(\!\boldsymbol{F}_{\rm R}\big((1\!-\!\beta)^2\sum_{l=1}^{K}\boldsymbol{f}_{{\rm B},l}\boldsymbol{f}_{{\rm B},l}^{\rm H}\!+\!\boldsymbol{R}_{\rm q}\big)\boldsymbol{F}_{\rm R}^{\rm H}\Big)\!\boldsymbol{h}_{\rm w}\!\bigg)\!\nonumber\\
		&=\frac{\sqrt{T}}{2\sigma_{\rm w}^2}{\rm Tr}\!\bigg(\!\Big(\!\boldsymbol{F}_{\rm R}\big((1\!-\!\beta)^2\sum_{l=1}^{K}\boldsymbol{f}_{{\rm B},l}\boldsymbol{f}_{{\rm B},l}^{\rm H}\!+\!\boldsymbol{R}_{\rm q}\big)\boldsymbol{F}_{\rm R}^{\rm H}\Big)\!\boldsymbol{\Omega}_{\rm w}\!\bigg)\!.
	\end{align}
	Note that (a) in \eqref{Ehw} is achieved by using the fact that $\xi - \ln(1+\xi) \leq \frac{\xi^2}{2},\forall \xi \geq 0$. Therefore, the transformed covertness constraint can be expressed as
	\begin{equation}\label{covertcstr}
		{\rm Tr}\!\bigg(\!\Big(\boldsymbol{F}_{\rm R}\big((1-\beta)^2\sum_{l=1}^{K}\boldsymbol{f}_{{\rm B},l}\boldsymbol{f}_{{\rm B},l}^{\rm H}\!+\!\boldsymbol{R}_{\rm q}\big)\boldsymbol{F}_{\rm R}^{\rm H}\Big)\boldsymbol{\Omega}_{\rm w}\!\bigg)\! \!\leq\! \frac{2\epsilon\sigma_{\rm w}^2}{\sqrt{T}}.
\end{equation}

Therefore, the optimization problem can be expressed as
\begin{subequations}\label{goalformulation_transformed}
	\begin{align}
		\underset{\boldsymbol{F}_{\rm R},\boldsymbol{F}_{\rm B}}{\max}\  & R_{\rm sum} \\
		\mathrm{s.t.\ } ~~&\eqref{goalformulation:sub2}, \eqref{powercstr}, \eqref{covertcstr}. 
	\end{align}
\end{subequations}
It is seen that the objective is expressed as a sum of logarithmic functions of fractional items. To solve the non-convexity of the objective, we employ the Lagrangian dual and quadratic transform \cite{Shen2018quadratic} and introduce auxiliary variables $\boldsymbol{r} \triangleq [r_1,r_2,\cdots,r_K]^{\rm T}$ and $\boldsymbol{z} \triangleq [z_1,z_2,\cdots,z_K]^{\rm T}$ so that $R_{\rm sum}$ can be replaced by
\begin{align}
	&f_{\rm r}(\boldsymbol{F}_{\rm R},\boldsymbol{F}_{\rm B},\boldsymbol{r},\boldsymbol{z}) \nonumber\\
	&\triangleq \sum_{k=1}^K\!\Big(\log(1\!+\!r_k)\!-\!r_k\!+\!2(1\!-\!\beta)\sqrt{1\!+\!r_k}{\rm Re}(z_k^{*}\boldsymbol{h}_k^{\rm H}\boldsymbol{F}_{\rm R}\boldsymbol{f}_{{\rm B},k})\nonumber\\
	&~~\!-\!|z_k|^2\big((1\!-\!\beta)^2\sum_{l=1}^{K}|\boldsymbol{h}_k^{\rm H}\boldsymbol{F}_{\rm R}\boldsymbol{f}_{{\rm B},l}|^2\!+\!\boldsymbol{h}_k^{\rm H}\boldsymbol{F}_{\rm R}\boldsymbol{R}_{\rm q}\boldsymbol{F}_{\rm R}^{\rm H}\boldsymbol{h}_k\!+\!\sigma_k^2\big)\Big).
\end{align}
Therefore, \eqref{goalformulation_transformed} is equivalent to
\begin{subequations}\label{fr}
	\begin{align}
		\underset{\boldsymbol{F}_{\rm R},\boldsymbol{F}_{\rm B},\boldsymbol{r},\boldsymbol{z}}{\max}\  & f_{\rm r}(\boldsymbol{F}_{\rm R},\boldsymbol{F}_{\rm B},\boldsymbol{r},\boldsymbol{z})\\
		\mathrm{s.t.\ } ~~~~&\eqref{goalformulation:sub2}, \eqref{powercstr}, \eqref{covertcstr}. 
	\end{align}
\end{subequations}

\begin{figure*}[ht]
	\vspace*{5pt}
	\setcounter{equation}{29}
	\begin{equation}\label{z_opt}
		\widehat{z}_{k} = \frac{(1-\beta)\sqrt{1+r_k}\boldsymbol{h}_k^{\rm H}\boldsymbol{F}_{\rm R}\boldsymbol{f}_{{\rm B,}k}}{(1-\beta)^2\sum_{l=1}^{K}|\boldsymbol{h}_k^{\rm H}\boldsymbol{F}_{\rm R}\boldsymbol{f}_{{\rm B},l}|^2+\boldsymbol{h}_k^{\rm H}\boldsymbol{F}_{\rm R}\boldsymbol{R}_{\rm q}\boldsymbol{F}_{\rm R}^{\rm H}\boldsymbol{h}_k+\sigma_k^2}, k = 1,2,\cdots,K.
	\end{equation}
	\rule[5pt]{18.3cm}{0.06em}
\end{figure*}

\vspace{-20pt}
\subsection{AO-based Hybrid Beamforming Design}\label{3_B}
Since it is difficult to optimize $\boldsymbol{F}_{\rm R},\boldsymbol{F}_{\rm B},\boldsymbol{r},\boldsymbol{z}$ simultaneously, we resort to the AO scheme~\cite{Qi2022Hybridswitches} to solve \eqref{fr}. The detailed procedures are presented as follows.

\emph{1) Optimization for $\boldsymbol{r}$:} When fixing $\boldsymbol{F}_{\rm R},\boldsymbol{F}_{\rm B},\boldsymbol{z}$, it can be seen $f_{\rm r}$ is a concave function of $\boldsymbol{r}$, thus the optimal $\widehat{r}_k$ can be obtained by letting $\partial f_{\rm r} / \partial r_k = 0$ for $k = 1,2,\cdots,K$, which can be expressed just as the form of ${\rm SIQNR}_k$ in \eqref{SIQNR}, i.e., 
\setcounter{equation}{28}
\begin{equation}\label{r_opt}
	\widehat{r}_{k} = {\rm SIQNR}_k, k = 1,2,\cdots,K.
\end{equation}

\emph{2) Optimization for $\boldsymbol{z}$:} When $\boldsymbol{F}_{\rm R},\boldsymbol{F}_{\rm B},\boldsymbol{r}$ are fixed, $f_{\rm r}$ is also a concave function of $\boldsymbol{z}$. Similarly, we let $\partial f_{\rm r} / \partial z_k = 0$ for $k = 1,2,\cdots,K$ and the optimal $\widehat{z}_{k}$ can be obtained as \eqref{z_opt} at the top of this page.

\emph{3) Optimization for $\boldsymbol{F}_{\rm R}$:} Given $\boldsymbol{F}_{\rm B},\boldsymbol{r},\boldsymbol{z}$, the optimization problem for $\boldsymbol{F}_{\rm R}$ can be expressed as
\setcounter{equation}{30}
\begin{subequations}\label{FRobt}
	\begin{align}
		\underset{\boldsymbol{F}_{\rm R}}{\max}\  &\sum_{k=1}^{K}\Big(2(1-\beta)\sqrt{1+r_k}{\rm Re}(z_k^*\boldsymbol{h}_k^{\rm H}\boldsymbol{F}_{\rm R}\boldsymbol{f}_{{\rm B,}k}) \nonumber \\
		&-|z_k|^2\big((1-\beta)^2\sum_{l=1}^{K}|\boldsymbol{h}_k^{\rm H}\boldsymbol{F}_{\rm R}\boldsymbol{f}_{{\rm B,}l}|^2+\boldsymbol{h}_k^{\rm H}\boldsymbol{F}_{\rm R}\boldsymbol{R}_{\rm q}\boldsymbol{F}_{\rm R}^{\rm H}\boldsymbol{h}_k\big)\Big) \label{FRobj}\\
		\mathrm{s.t.\ } ~&\eqref{goalformulation:sub2}, \eqref{powercstr}, \eqref{covertcstr}.
	\end{align}
\end{subequations}

The objective function in \eqref{FRobt} related to $\boldsymbol{F}_{\rm R}$ can be transformed as
\begin{align}\label{gr_fr}
	&\sum_{k=1}^{K}\Big(2(1-\beta)\sqrt{1+r_k}{\rm Re}(z_k^*\boldsymbol{h}_k^{\rm H}\boldsymbol{F}_{\rm R}\boldsymbol{f}_{{\rm B,}k}) \nonumber \\
	&~~~-|z_k|^2\big((1-\beta)^2\sum_{l=1}^{K}|\boldsymbol{h}_k^{\rm H}\boldsymbol{F}_{\rm R}\boldsymbol{f}_{{\rm B,}l}|^2+\boldsymbol{h}_k^{\rm H}\boldsymbol{F}_{\rm R}\boldsymbol{R}_{\rm q}\boldsymbol{F}_{\rm R}^{\rm H}\boldsymbol{h}_k\big)\Big) \nonumber\\
	&\overset{(a)}{=}\sum_{k=1}^{K}\bigg(2(1-\beta)\sqrt{1+r_k}{\rm Re}\big(z_k^*{\rm Tr}(\boldsymbol{f}_{{\rm B,}k}\boldsymbol{h}_k^{\rm H}\boldsymbol{F}_{\rm R})\big) \nonumber \\
	&~~~-|z_k|^2\bigg({\rm Tr}\Big(\boldsymbol{F}_{\rm R}\big((1-\beta)^2\boldsymbol{F}_{\rm B}\boldsymbol{F}_{\rm B}^{\rm H}+\boldsymbol{R}_{\rm q}\big)\boldsymbol{F}_{\rm R}^{\rm H}\boldsymbol{h}_{k}\boldsymbol{h}_k^{\rm H}\Big)\bigg)\Bigg) \nonumber\\
	&\overset{(b)}{=}-\boldsymbol{f}_{\rm R}^{\rm H}\boldsymbol{Q}_0\boldsymbol{f}_{\rm R}-2{\rm Re}(\boldsymbol{p}_0^{\rm H}\boldsymbol{f}_{\rm R}),
\end{align}
where
\begin{align}
	&\boldsymbol{f}_{\rm R} \triangleq {\rm vec}(\boldsymbol{F}_{\rm R}), \\
	&\boldsymbol{Q}_0 \triangleq \sum_{k=1}^K |z_k|^2\Big(\big((1-\beta)^2(\boldsymbol{F}_{\rm B}\boldsymbol{F}_{\rm B}^{\rm H})^{\rm T}+\boldsymbol{R}_{\rm q}^{\rm T}\big)\otimes (\boldsymbol{h}_k\boldsymbol{h}_k^{\rm H})\Big), \\
	&\boldsymbol{p}_0 \triangleq -\sum_{k=1}^K (1-\beta)z_k\sqrt{1+r_k}(\boldsymbol{I}_{K}\otimes \boldsymbol{h}_k)\boldsymbol{f}_{{\rm B,}k}^*.
\end{align}
Note that the equality (a) in \eqref{gr_fr} holds because ${\rm Tr}(\boldsymbol{A}\boldsymbol{B}) = {\rm Tr}(\boldsymbol{B}\boldsymbol{A})$. The equality (b) holds due to the fact that ${\rm Tr}(\boldsymbol{A}^{\rm H}\boldsymbol{B}\boldsymbol{C})={\rm vec}(\boldsymbol{A})^{\rm H}(\boldsymbol{I} \otimes \boldsymbol{B}){\rm vec}(\boldsymbol{C})$ and ${\rm Tr}(\boldsymbol{A}\boldsymbol{B}\boldsymbol{A}^{\rm H}\boldsymbol{C})={\rm vec}(\boldsymbol{A})^{\rm H}(\boldsymbol{B}^{\rm T}\otimes \boldsymbol{C}){\rm vec}(\boldsymbol{A})$.

Similarly, we can rewrite the constraints \eqref{powercstr} and \eqref{covertcstr}. Then, \eqref{FRobt} can be equivalently expressed as
\begin{subequations}\label{fRfinalopt}
	\begin{align}
		\underset{\boldsymbol{f}_{\rm R}}{\min}\ &g_{\rm r}(\boldsymbol{f}_{\rm R})\triangleq\boldsymbol{f}_{\rm R}^{\rm H}\boldsymbol{Q}_0\boldsymbol{f}_{\rm R}+2{\rm Re}(\boldsymbol{p}_0^{\rm H}\boldsymbol{f}_{\rm R}) \label{fRfinalopt:sub1}\\
		\mathrm{s.t.\ } ~&\boldsymbol{f}_{\rm R}^{\rm H}\boldsymbol{Q}_1\boldsymbol{f}_{\rm R} \leq P_{\rm max}, \label{fRfinalopt:sub2}\\
		~&\boldsymbol{f}_{\rm R}^{\rm H}\boldsymbol{Q}_2\boldsymbol{f}_{\rm R} \leq \frac{2\epsilon \sigma_{\rm w}^2}{\sqrt{T}}, \label{fRfinalopt:sub3}\\
		~&\big|[\boldsymbol{f}_{\rm R}]_u\big| = 1, u=1,2,\cdots,KN, \label{fRfinalopt:sub4}
	\end{align}
\end{subequations}
where 
\begin{align}
	&\boldsymbol{Q}_1 \triangleq \big((1-\beta)^2(\boldsymbol{F}_{\rm B}\boldsymbol{F}_{\rm B}^{\rm H})^{\rm T}\big) \otimes \boldsymbol{I}_{N},\\
	&\boldsymbol{Q}_2 \triangleq \big((1-\beta)^2(\boldsymbol{F}_{\rm B}\boldsymbol{F}_{\rm B}^{\rm H})^{\rm T}\big) \otimes \boldsymbol{\Omega}_{\rm w}.
\end{align}
The newly obtained \eqref{fRfinalopt:sub2}, \eqref{fRfinalopt:sub3} and $\eqref{fRfinalopt:sub4}$ correspond to the original \eqref{powercstr}, \eqref{covertcstr} and \eqref{goalformulation:sub2}, respectively. Therefore, the problem in~\eqref{fRfinalopt} is transformed into a standard QCQP problem with additional constant-modulus constraints, enabling us to solve it efficiently with the iMM method~\cite{He2022QCQP}.

We start by introducing the majorization-minimization method~\cite{Huang2022PTCISAC}, which can find a majorized function for the quadratic term under the constant-modulus constraints. For example, we assume that $\boldsymbol{x} \in \mathbb{C}^N$ has the constant-modulus constraints, i.e., $\big|[\boldsymbol{x}]_u\big|=1$ for $u= 1,2,\cdots,N$ and $\boldsymbol{Q}$ is a Hermitian matrix. Once we obtain a feasible solution $\overline{\boldsymbol{x}}$ satisfying the constant-modulus constraints, the quadratic term $\boldsymbol{x}^{\rm H}\boldsymbol{Q}\boldsymbol{x}$ can be upper-bounded by
\begin{equation}\label{MMinequaility}
	\boldsymbol{x}^{\rm H}\boldsymbol{Q}\boldsymbol{x} \leq 2\lambda N +2{\rm Re}\big(\boldsymbol{x}^{\rm H}(\boldsymbol{Q}-\lambda\boldsymbol{I}_N)\overline{\boldsymbol{x}}\big) - (\overline{\boldsymbol{x}})^{\rm H}\boldsymbol{Q}\overline{\boldsymbol{x}},
\end{equation}
where $\lambda$ is usually selected as the maximum eigenvalue or the trace of $\boldsymbol{Q}$ to satisfy $\lambda \boldsymbol{I}_N - \boldsymbol{Q} \succeq \boldsymbol{0}$. It is seen that the right hand of \eqref{MMinequaility} is a linear form of $\boldsymbol{x}$ which facilitates the optimization. Therefore, once we obtain a feasible solution $\overline{{\boldsymbol{f}}}_{\rm R}$, we can derive a surrogate problem for \eqref{fRfinalopt} as
\begin{subequations}\label{fRfinalopt_upper}
	\begin{align}
		\underset{\boldsymbol{f}_{\rm R}}{\min}\ & g_0(\boldsymbol{f}_{\rm R}) \\
		\mathrm{s.t.\ } ~& g_v(\boldsymbol{f}_{\rm R}) \leq 0, v = 1,2, \\
		~&\big|[\boldsymbol{f}_{\rm R}]_u\big| = 1, u=1,2,\cdots,KN, \label{fRfinalopt_upper:sub3}
	\end{align}
\end{subequations}
where we define
\begin{align}
	g_0(\boldsymbol{f}_{\rm R})& \triangleq 2{\rm Re}\Big(\boldsymbol{f}_{\rm R}^{\rm H}\big((\boldsymbol{Q}_0-\lambda_0\boldsymbol{I}_{KN})\overline{{\boldsymbol{f}}}_{\rm R}+\boldsymbol{p}_0\big)\Big)\\
	g_1(\boldsymbol{f}_{\rm R}) &\triangleq 2{\rm Re}\Big(\boldsymbol{f}_{\rm R}^{\rm H}\big((\boldsymbol{Q}_1-\lambda_1\boldsymbol{I}_{KN})\overline{{\boldsymbol{f}}}_{\rm R}\big)\Big)+2\lambda_1 KN\nonumber\\
	&~~~- \overline{\boldsymbol{f}}_{\rm R}^{\rm H}\boldsymbol{Q}_1\overline{\boldsymbol{f}}_{\rm R} - P_{\rm max},\label{g_1}\\
	g_2(\boldsymbol{f}_{\rm R}) &\triangleq 2{\rm Re}\Big(\boldsymbol{f}_{\rm R}^{\rm H}\big((\boldsymbol{Q}_2-\lambda_2\boldsymbol{I}_{KN})\overline{{\boldsymbol{f}}}_{\rm R}\big)\Big)+2\lambda_2 KN\nonumber\\
	&~~~- \overline{\boldsymbol{f}}_{\rm R}^{\rm H}\boldsymbol{Q}_2\overline{\boldsymbol{f}}_{\rm R}- \frac{2\epsilon\sigma_{\rm w}^2}{\sqrt{T}}, \label{g_2}
\end{align}
and $\lambda_v$ denotes the trace of $\boldsymbol{Q}_v$ for $v = 0,1,2$.

Since \eqref{fRfinalopt_upper} is still non-convex due to \eqref{fRfinalopt_upper:sub3}, we can solve its Lagrange dual problem which is expressed as
\begin{equation}\label{LDP}
	\underset{\{\omega_1,\omega_2 \geq 0\}}{\sup}\underset{\{|[\boldsymbol{f}_{\rm R}]_u| = 1\}_{u=1}^{KN}}{\min} g_0(\boldsymbol{f}_{\rm R}) + \sum_{v=1}^2 \omega_v g_v(\boldsymbol{f}_{\rm R}).
\end{equation}
As the objective in \eqref{LDP} is a linear function of $\boldsymbol{f}_{\rm R}$, the optimal solution for $\boldsymbol{f}_{\rm R}$ can be given by
\begin{align}\label{fRsolution}
	&\widehat{\boldsymbol{f}}_{\rm R}(\omega_1,\omega_2) = {\rm exp}\bigg(j \angle\Big(\sum_{v=1}^2 \big((\lambda_v \boldsymbol{I}_{KN}-\boldsymbol{Q}_v)\overline{\boldsymbol{f}}_{\rm R}\big)\omega_v \nonumber\\
	&~~~~~~~~~~+ (\lambda_0 \boldsymbol{I}_{KN}-\boldsymbol{Q}_0)\overline{\boldsymbol{f}}_{\rm R}\Big)\bigg). 
\end{align}
Additionally, we consider the remaining Karush-Kuhn-Tucker conditions, i.e.,
\begin{align}
	&g_v\big(\widehat{\boldsymbol{f}}_{\rm R}(\omega_1,\omega_2)\big) \leq 0, \omega_v \geq 0, v = 1,2, \\
	&\omega_v g_v\big(\widehat{\boldsymbol{f}}_{\rm R}(\omega_1,\omega_2)\big) = 0, v = 1,2.
\end{align}					
Based on ~\cite[Lemma 3]{He2022QCQP}, the  bisection method can be used to obtain the numerical results for $\omega_1$ and $\omega_2$. Specifically, when $\omega_2$ is fixed, we use the bisection method to search $\hat{\omega}_1$ which satisfies $g_1\big(\widehat{\boldsymbol{f}}_{\rm R}(\hat{\omega}_1,\omega_2)\big) = 0$. Similar techniques are used to obtain $\hat{\omega_2}$. We can alternately optimize the dual variables $\omega_1$ and $\omega_2$ until convergence. By using the converged dual variables, we can derive the optimal $\boldsymbol{f}_{\rm R}$ via \eqref{fRsolution} Then, we define $\overline{g}_v(a) \triangleq g_v\big(\widehat{\boldsymbol{f}}_{\rm R}(\omega_1,\omega_2)\big)\big|_{\omega_v=a}$ for $v = 1,2$ for simplicity and the iMM-based analog beamforming design is summarized in \textbf{Algorithm}~\ref{alg1}.

\begin{algorithm}[!t]
	\caption{iMM-based Analog Beamforming Design}
	\label{alg1}
	\begin{algorithmic}[1]
			\STATE \textbf{Input}: A feasible solution $\overline{\boldsymbol{f}}_{\rm R}^{(0)}$ for \eqref{fRfinalopt}, the convergence thresholds $\varepsilon_1$, $\varepsilon_2$, the solution accuracy $\delta$, the iteration counter $i_1 \leftarrow 0$ and $\boldsymbol{\omega}^{(0)} \leftarrow [0,0]$.
			\REPEAT
			\STATE Update $g_1$ and $g_2$ by replacing $\overline{\boldsymbol{f}}_{\rm R}$ with $\overline{\boldsymbol{f}}_{\rm R}^{(i_1)}$ in \eqref{g_1} and \eqref{g_2}, respectively. 
			\STATE Set $i_2 \leftarrow 0$.
			\REPEAT
			\FOR{$v = 1,2$}
			\IF{$\overline{g}_v(0) < 0$}
			\STATE $\hat{\omega}_v \leftarrow 0$.
			\ELSIF{$\big|\lim\limits_{\omega_v\to +\infty}\overline{g}_v(\omega_v)\big|\leq \delta$}
			\STATE $\hat{\omega}_v \leftarrow +\infty$.
			\STATE Update $\boldsymbol{f}_{\rm R}$ via \eqref{fRsolution}. Go to step 22.
			\ELSE
			\STATE Obtain $\hat{x}$ by solving $\overline{g}_v(\hat{x}) = 0, \hat{x} \geq 0$ via the bisection method.
			\STATE Update $\hat{\omega}_v \leftarrow \hat{x}$.
			\ENDIF
			\ENDFOR
			\STATE Set $i_2 \leftarrow i_2 + 1$ and update $\boldsymbol{\omega}^{(i_2)} \leftarrow [\hat{\omega}_1,\hat{\omega}_2]$.
			\UNTIL{$\|\boldsymbol{\omega}^{(i_2)}-\boldsymbol{\omega}^{(i_2-1)}\|_2 \leq \varepsilon_2$}
			\STATE Set $\boldsymbol{\omega}^{(0)} \leftarrow \boldsymbol{\omega}^{(i_2)}$, $i_1 \leftarrow i_1 + 1$ and update $\overline{\boldsymbol{f}}_{\rm R}^{(i_1)}$ via \eqref{fRsolution}.
			\UNTIL{$|g_{\rm r}(\overline{\boldsymbol{f}}_{\rm R}^{(i_1-1)}) - g_{\rm r}(\overline{\boldsymbol{f}}_{\rm R}^{(i_1)})|/|g_{\rm r}(\overline{\boldsymbol{f}}_{\rm R}^{(i_1-1)})| \leq \varepsilon_1$.}
			\STATE Set $\boldsymbol{f}_{\rm R} \leftarrow \overline{\boldsymbol{f}}_{\rm R}^{(i_1)}$.
			\STATE \textbf{Output}: $\boldsymbol{f}_{\rm R}$.
		\end{algorithmic}
\end{algorithm}

\emph{4) Optimization for $\boldsymbol{F}_{\rm B}$:} Given $\boldsymbol{F}_{\rm R},\boldsymbol{r},\boldsymbol{z}$, the objective function in \eqref{fr} related to $\boldsymbol{F}_{\rm B}$ can be expressed as
\begin{align}\label{objforfB}
	&\sum_{k=1}^{K}\Big(2(1-\beta)\sqrt{1+r_k}{\rm Re}(z_k^*\boldsymbol{h}_k^{\rm H}\boldsymbol{F}_{\rm R}\boldsymbol{f}_{{\rm B,}k}) \nonumber \\
	&~~~-|z_k|^2\big((1-\beta)^2\sum_{l=1}^{K}|\boldsymbol{h}_k^{\rm H}\boldsymbol{F}_{\rm R}\boldsymbol{f}_{{\rm B,}l}|^2+\boldsymbol{h}_k^{\rm H}\boldsymbol{F}_{\rm R}\boldsymbol{R}_{\rm q}\boldsymbol{F}_{\rm R}^{\rm H}\boldsymbol{h}_k\big)\Big) \nonumber\\
	&\overset{(a)}{=} \sum_{k=1}^{K} \Big(2(1-\beta)\sqrt{1+r_k}{\rm Re}(z_k^*\boldsymbol{h}_k^{\rm H}\boldsymbol{F}_{\rm R}\boldsymbol{f}_{{\rm B,}k}) \nonumber \\
	&~~~-|z_k|^2 \sum_{l=1}^{K}\big((1-\beta)^2\boldsymbol{f}_{\rm B}^{\rm H}\boldsymbol{S}_l^{\rm H}\boldsymbol{F}_{\rm R}^{\rm H}\boldsymbol{h}_k\boldsymbol{h}_k^{\rm H}\boldsymbol{F}_{\rm R}\boldsymbol{S}_l\boldsymbol{f}_{\rm B} \nonumber \\
	&~~~+\beta(1-\beta)\boldsymbol{f}_{\rm B}^{\rm H}\boldsymbol{S}_l^{\rm H}{\rm diag}(\boldsymbol{F}_{\rm R}^{\rm H}\boldsymbol{h}_k\boldsymbol{h}_k^{\rm H}\boldsymbol{F}_{\rm R})\boldsymbol{S}_l\boldsymbol{f}_{\rm B}\big)\Big) \nonumber\\
	& = -\boldsymbol{f}_{\rm B}^{\rm H} \boldsymbol{\Xi}_0 \boldsymbol{f}_{\rm B} - 2{\rm Re}(\boldsymbol{\phi}_0^{\rm H}\boldsymbol{f}_{\rm B}),
\end{align} 
where
\begin{align}
	&\boldsymbol{f}_{\rm B} \triangleq {\rm vec}(\boldsymbol{F}_{\rm B}), \\
	&\boldsymbol{S}_l \triangleq [\overbrace{\boldsymbol{0},\cdots,\boldsymbol{0}}^{l-1},\boldsymbol{I}_{K},\overbrace{\boldsymbol{0},\cdots,\boldsymbol{0}}^{K-l}], l = 1,2,\cdots,K, \\
	&\boldsymbol{\Xi}_0 \triangleq \sum_{k=1}^{K} |z_k|^2 \Big(\sum_{l=1}^{K}\big((1-\beta)^2\boldsymbol{S}_l^{\rm H}\boldsymbol{F}_{\rm R}^{\rm H}\boldsymbol{h}_k\boldsymbol{h}_k^{\rm H}\boldsymbol{F}_{\rm R}\boldsymbol{S}_l\nonumber\\
	&~~~~~~+\beta(1-\beta)\boldsymbol{S}_l^{\rm H}{\rm diag}(\boldsymbol{F}_{\rm R}^{\rm H}\boldsymbol{h}_k\boldsymbol{h}_k^{\rm H}\boldsymbol{F}_{\rm R})\boldsymbol{S}_l\big)\Big), \\
	&\boldsymbol{\phi}_0 \triangleq -\sum_{k=1}^{K}\big((1-\beta)z_k\sqrt{1+r_k} \boldsymbol{S}_k^{\rm H}\boldsymbol{F}_{\rm R}^{\rm H}\boldsymbol{h}_k\big).
\end{align}
Note that (a) in \eqref{objforfB} holds due to the facts that $\boldsymbol{S}_k\boldsymbol{f}_{\rm B} = \boldsymbol{f}_{{\rm B,}k}$ for $k = 1,2,\cdots,K$ and ${\rm Tr}\big(\boldsymbol{A}{\rm diag}(\boldsymbol{B})\big) = {\rm Tr}\big({\rm diag}(\boldsymbol{A})\boldsymbol{B}\big)$. 

Similar operations can be performed for the constraints \eqref{powercstr} and \eqref{covertcstr}, then we can derive the optimization problem for $\boldsymbol{f}_{\rm B}$ as
\begin{subequations}\label{fBfinalopt}
	\begin{align}
		\min_{\boldsymbol{f}_{\rm B}}\ &\boldsymbol{f}_{\rm B}^{\rm H} \boldsymbol{\Xi}_0 \boldsymbol{f}_{\rm B} + 2{\rm Re}(\boldsymbol{\phi}_0^{\rm H}\boldsymbol{f}_{\rm B}) \label{fBfinalopt:sub1}\\
		\mathrm{s.t.\ } ~&\boldsymbol{f}_{\rm B}^{\rm H}\boldsymbol{\Xi}_1\boldsymbol{f}_{\rm B} \leq P_{\rm max}, \label{fBfinalopt:sub2}\\
		~&\boldsymbol{f}_{\rm B}^{\rm H}\boldsymbol{\Xi}_2\boldsymbol{f}_{\rm B} \leq \frac{2\epsilon \sigma_{\rm w}^2}{\sqrt{T}}, \label{fBfinalopt:sub3}
	\end{align}
\end{subequations}
where we define
\begin{align}
	\boldsymbol{\Xi}_1 \triangleq \sum_{k=1}^K \boldsymbol{S}_k^{\rm H}\big((1-\beta&)^2\boldsymbol{F}_{\rm R}^{\rm H}\boldsymbol{F}_{\rm R} \nonumber\\
	&+\beta(1-\beta){\rm diag}(\boldsymbol{F}_{\rm R}^{\rm H}\boldsymbol{F}_{\rm R})\big)\boldsymbol{S}_k,
\end{align}
\begin{align}
	\boldsymbol{\Xi}_2 \triangleq \sum_{k=1}^K\boldsymbol{S}_k^{\rm H}\big((1-\beta&)^2\boldsymbol{F}_{\rm R}^{\rm H}\boldsymbol{\Omega}_{\rm w}\boldsymbol{F}_{\rm R}\nonumber\\
	&+\beta(1-\beta){\rm diag}(\boldsymbol{F}_{\rm R}^{\rm H}\boldsymbol{\Omega}_{\rm w}\boldsymbol{F}_{\rm R})\big)\boldsymbol{S}_k\big).
\end{align}

Since that $\boldsymbol{\Xi}_0,\boldsymbol{\Xi}_1$ and $\boldsymbol{\Xi}_2$ are all Hermitian semi-definite matrices, \eqref{fBfinalopt} is a standard QCQP convex problem, which can be solved by existing solvers (e.g., Mosek optimization tools) with the interior-point method. 

With an appropriate initialization for $\boldsymbol{F}_{\rm R}$ and $\boldsymbol{F}_{\rm B}$, we can alternately optimize $\boldsymbol{r}$, $\boldsymbol{z}$, $\boldsymbol{F}_{\rm R}$ and $\boldsymbol{F}_{\rm B}$ until convergence. Finally, we summarize the proposed AO scheme for hybrid beamforming design in \textbf{Algorithm}~\ref{alg2}. 

\begin{algorithm}[!t]
	\caption{AO-based Hybrid Beamforming Design}
	\label{alg2}
	\begin{algorithmic}[1]
	\STATE \textbf{Input}: $\boldsymbol{h}_1,\cdots,\boldsymbol{h}_K$, $\boldsymbol{\Omega}_{\rm w}$, $P_{\rm max}$, $\epsilon$.
	\STATE Initialize $\boldsymbol{F}_{\rm R}$, $\boldsymbol{F}_{\rm B}$.
	\REPEAT
	\STATE Update $r_k, k = 1,2,\cdots,K$ via \eqref{r_opt}.
	\STATE Update $z_k, k = 1,2,\cdots,K$ via \eqref{z_opt}.
	\STATE Update analog beamformer $\boldsymbol{f}_{\rm R}$ with \textbf{Algorithm}~\ref{alg1}.
	\STATE Update digital beamformer $\boldsymbol{f}_{\rm B}$ by solving \eqref{fBfinalopt}.
	\UNTIL{$f_{\rm r}$ in \eqref{fr} is converged.}
	\STATE \textbf{Output}: $\boldsymbol{F}_{\rm R} \leftarrow {\rm vec}^{-1}(\boldsymbol{f}_{\rm R})$, $\boldsymbol{F}_{\rm B} \leftarrow {\rm vec}^{-1}(\boldsymbol{f}_{\rm B})$.
	\end{algorithmic}
\end{algorithm}

\subsection{Complexity Analysis}\label{3_C}
In this subsection, we analyze the computational complexity of the proposed AO scheme in \textbf{Algorithm}~\ref{alg2}. In fact, steps 2, 4, 5, 6 and 7 take up dominant computational cost. Specifically, in step 2, the computational complexity is denoted as $\mathcal{O}_{\rm Init}$ which varies with different initialization approach. Computing the auxiliary variables $\boldsymbol{r}$ and $\boldsymbol{z}$ involves the same order of complexity as $\mathcal{O}(N K^2)$ in step 4 and 5, respectively. Based on~\cite{He2022QCQP}, the complexity of \textbf{Algorithm}~\ref{alg1} in step 6 is $\mathcal{O}\big(N_{\rm max}^{(1)}(N K)^2\big)$, where $N_{\rm max}^{(1)}$ denotes the total number of iterations in \textbf{Algorithm}~\ref{alg1}. Finally, in step 7, solving the QCQP convex problem for $\boldsymbol{f}_{\rm B}$ with an interior-point method involves the complexity $\mathcal{O}\big(K^{7}\big)$. Therefore, the overall complexity of the proposed AO scheme is approximately $\mathcal{O}_{\rm Init} + \mathcal{O}\Big(N_{\rm max}^{(2)}\big((N_{\rm max}^{(1)}(N K)^2+K^{7}\big)\Big)$, where $N_{\rm max}^{(2)}$ denotes the number of the outer iterations in \textbf{Algorithm}~\ref{alg2}.

\section{VSH scheme for Hybrid Beamforming Design}\label{section4}
To reduce the computational complexity of the AO scheme, we propose a VSH scheme for hybrid beamforming design, which can also be used to obtain an initialization for the AO scheme to further improve its performance. Subsequently, we prove that the SCR obtained by the VSH scheme can achieve the channel mutual information as the number of antennas grows to infinity. To further enhance the SCR performance, we formulate a novel power allocation problem and solve it using an alternating method with closed-form solutions.

\subsection{VSH-based Hybrid Beamforming Design}\label{S4subA}
We first combine the received signals in \eqref{ykwithQ} from all the $K$ users and denote $\boldsymbol{y} = [y_1,y_2,\cdots,y_K]^{\rm T}$, $\boldsymbol{s} = [s_1,s_2,\cdots,s_K]^{\rm T}$, $\boldsymbol{H} = [\boldsymbol{h}_1,\boldsymbol{h}_2,\cdots,\boldsymbol{h}_K]^{\rm H}$ and $\overline{\boldsymbol{\eta}} = [\overline{\eta}_1,\overline{\eta}_2,\cdots,\overline{\eta}_K]^{\rm T}$. Then, the received signal vector $\boldsymbol{y}$ can be expressed as
\begin{equation}\label{alluser}
	\boldsymbol{y} = (1-\beta)\boldsymbol{H}\boldsymbol{F}_{\rm R}\boldsymbol{F}_{\rm B}\boldsymbol{s} + \boldsymbol{H}\boldsymbol{F}_{\rm R}\boldsymbol{\eta}_{\rm q} + \overline{\boldsymbol{\eta}}.
\end{equation}
By assuming that all the transmitted symbols obey independently circularly symmetric complex Gaussian distribution, the mutual information between $\boldsymbol{s}$ and $\boldsymbol{y}$ can be defined as
\begin{equation}\label{MI}
	\mathcal{I}(\boldsymbol{s};\boldsymbol{y}) \triangleq \log\Big|\boldsymbol{I}_K + (\boldsymbol{\Upsilon})^{-1}\big((1-\beta)^2 \boldsymbol{H}\boldsymbol{F}_{\rm R}\boldsymbol{F}_{\rm B}\boldsymbol{F}_{\rm B}^{\rm H}\boldsymbol{F}_{\rm R}^{\rm H}\boldsymbol{H}^{\rm H}\big)\Big|,
\end{equation}
where
\begin{equation}\label{Intferenosie}
	\boldsymbol{\Upsilon} \triangleq \boldsymbol{R}_{\overline{\eta}} + \boldsymbol{H}\boldsymbol{F}_{\rm R}\boldsymbol{R}_{\rm q}\boldsymbol{F}_{\rm R}^{\rm H}\boldsymbol{H}^{\rm H}
\end{equation}
and $\boldsymbol{R}_{\overline{\eta}} \triangleq {\rm Diag}(\sigma_1^2,\sigma_2^2,\cdots,\sigma_K^2)$. In fact, the sum rate in \eqref{Rsum} is always less than or equal to the mutual information in \eqref{MI}~\cite{EIT2002}. Therefore, we aim to design hybrid beamformers so that the sum rate approximates the mutual information. Since a large number of antennas are equipped in the massive MIMO systems, we consider $N$ is approximately to be infinity in this section. 

For the optimization problem in \eqref{goalformulation_transformed}, we first consider the covertness constraint \eqref{covertcstr}. Due to the existence of the term $\boldsymbol{R}_{\rm q}$ caused by the quantization noise in \eqref{covertcstr}, we cannot employ the zero-forcing digital beamforming \cite{Peng2021Strategies,Ma2021Robust} straightforwardly. Fortunately, it can be seen that the left hand of \eqref{covertcstr} will be zero when $\boldsymbol{\Omega}_{\rm w}\boldsymbol{F}_{\rm R} = 0$, indicating that $\boldsymbol{F}_{\rm R}$ should lie in the null space of $\boldsymbol{\Omega}_{\rm w}$ to achieve the covertness. Although the analog beamformer has the constant-modulus constraints in \eqref{goalformulation:sub2}, we can eliminate these constraints by doubling the number of RF chains~\cite{Wu2018Hysbd}. Therefore, we temporally ignore the constant-modulus constraints and perform the singular value decomposition (SVD) of $\boldsymbol{\Omega}_{\rm w}$, i.e.,
\begin{equation}
	\boldsymbol{\Omega}_{\rm w} = \boldsymbol{U}_{\rm w}\boldsymbol{\Lambda}_{\rm w}[\boldsymbol{V}_{\rm w}^{(1)} \boldsymbol{V}_{\rm w}^{(0)}]^{\rm H}, 
\end{equation}
where $\boldsymbol{V}_{\rm w}^{(0)} \in \mathbb{C}^{N \times N_0}$, $N-D_{\rm w} \leq N_0 < N$ and $\boldsymbol{V}_{\rm w}^{(1)}\in \mathbb{C}^{N \times (N - N_0)}$ denote the null space and the column space of $\boldsymbol{\Omega}_{\rm w}$, respectively. Then we can define $\widehat{\boldsymbol{H}} \triangleq \boldsymbol{H}\boldsymbol{V}_{\rm w}^{(0)}$ and perform the SVD that $\widehat{\boldsymbol{H}} = \widehat{\boldsymbol{U}}\widehat{\boldsymbol{\Lambda}}\widehat{\boldsymbol{V}}^{\rm H}$. From the right singular matrix $\widehat{\boldsymbol{V}}$, we choose the $K$ strongest components denoted as $[\widehat{\boldsymbol{V}}]_{:,1:K}$ and $\boldsymbol{V}_{\rm w}^{(0)}$ to constitute the analog beamformer, which can be expressed as
\begin{equation}\label{FR_Heur}
	\widetilde{\boldsymbol{F}}_{\rm R} = \boldsymbol{V}_{\rm w}^{(0)} [\widehat{\boldsymbol{V}}]_{:,1:K}.
\end{equation}

Subsequently, we will use the block diagonalization method, which is effective to eliminate the multiuser interference, to design the digital beamformer $\boldsymbol{F}_{\rm B}$. We start by introducing $\widetilde{\boldsymbol{h}}_k = \boldsymbol{h}_{\rm k}^{\rm H}\boldsymbol{F}_{\rm R}$ and $\widetilde{\boldsymbol{H}}_k = [\widetilde{\boldsymbol{h}}_1^{\rm T},\cdots,\widetilde{\boldsymbol{h}}_{k-1}^{\rm T},\widetilde{\boldsymbol{h}}_{k+1}^{\rm T},\cdots,\widetilde{\boldsymbol{h}}_K^{\rm T}]^{\rm T}$. Then $\boldsymbol{f}_{{\rm B,}k}$ should lie in the null space of $\widetilde{\boldsymbol{H}}_k$. Similarly, we perform the SVD, i.e., $\widetilde{\boldsymbol{H}}_k = \widetilde{\boldsymbol{U}}_k\widetilde{\boldsymbol{\Lambda}}_k[\widetilde{\boldsymbol{V}}_k^{(1)} \widetilde{\boldsymbol{V}}_k^{(0)}]^{\rm H}$, where $\widetilde{\boldsymbol{V}}_k^{(0)} \in \mathbb{C}^{K}$ and $\widetilde{\boldsymbol{V}}_k^{(1)}\in \mathbb{C}^{K \times (K - 1)}$ denote the null space and the column space of $\widetilde{\boldsymbol{H}}_k$, respectively. Therefore, we let $\boldsymbol{f}_{{\rm B,}k} = \zeta_k \widetilde{\boldsymbol{V}}_k^{(0)}$ and the digital beamformer can be expressed as
\begin{equation}\label{FB_Heur}
	\widetilde{\boldsymbol{F}}_{\rm B} = [\widetilde{\boldsymbol{V}}_1^{(0)},\widetilde{\boldsymbol{V}}_2^{(0)},\cdots,\widetilde{\boldsymbol{V}}_K^{(0)}]\boldsymbol{\Sigma}_{\rm B},
\end{equation}
where $\boldsymbol{\Sigma}_{\rm B} \triangleq {\rm Diag}(\zeta_1,\zeta_2,\cdots,\zeta_K) \in \mathbb{C}^{K\times K}$ and we define $p_k \triangleq |\zeta_k|^2$ for $k = 1,2,\cdots,K$ as the power allocated to the $k$th data streams. Then, the following proposition is derived to validate the effectiveness of the aforementioned analog and digital beamforming design.

\emph{Proposition 2:} The analog and digital beamformers designed by the VSH scheme in \eqref{FR_Heur} and \eqref{FB_Heur} ensure that the SCR in~\eqref{Rsum} is equivalent to the mutual information in~\eqref{MI} when $N$ grows to be infinity.

\emph{Proof:} For any continuous distribution $\theta_{\rm w}^{(d_{\rm w})}$ for $d_{\rm w} = 0,1,\cdots,D_{\rm w}$ and $\theta_k^{(d_k)}$ for $d_k = 0,1,\cdots,D_k$, there exists ${\rm Pr}(\theta_{\rm w}^{(d_{\rm w})} = \theta_k^{(d_k)}) = 0$. Therefore, when $N$ grows to be infinity, we can obtain
\begin{equation}\label{AoDneq}
	{\rm Pr}\big(\lim\limits_{N\to \infty}\boldsymbol{a}(N,\theta_{\rm w}^{(d_{\rm w})})^{\rm H}\boldsymbol{a}(N,\theta_k^{(d_k)})\big) = 0.
\end{equation}
Then, we can further derive that
\begin{equation}\label{omgea_wH}
	{\rm Pr}(\lim\limits_{N\to \infty}\boldsymbol{\Omega}_{\rm w}\boldsymbol{H}^{\rm H}) = \boldsymbol{0}_K,
\end{equation} 
where $\boldsymbol{0}_K$ denotes a zero column vector with the length of $K$.

From \eqref{omgea_wH} we can find that $\boldsymbol{H}^{\rm H}$ is in the null space of $\boldsymbol{\Omega}_{\rm w}$. By performing the SVD of $\boldsymbol{H}$, i.e., $\boldsymbol{H} = \boldsymbol{U}\boldsymbol{\Lambda}\boldsymbol{V}^{\rm H}$, we can obtain that $[\boldsymbol{V}]_{:,1:K} \in \boldsymbol{V}_{\rm w}^{(0)}$, where $[\boldsymbol{V}]_{:,1:K}$ denotes the primary $K$ components of $\boldsymbol{V}$. Therefore, \eqref{FR_Heur} is equivalent to
\begin{equation}\label{FRequiv}
	\widetilde{\boldsymbol{F}}_{\rm R} = [\boldsymbol{V}]_{:,1:K}.
\end{equation}
Since ${\rm Pr}(\theta_l^{(d_l)} = \theta_k^{(d_k)}) = 0, \forall d_k \in \{1,2,\cdots, D_k\}, d_l \in \{1,2,\cdots, D_l\}, l \neq k$, we can obtain
\begin{equation}\label{hkhm}
	{\rm Pr}\big(\lim\limits_{N \to \infty} \boldsymbol{h}_k^{\rm H} \boldsymbol{h}_l\big) = 0, \forall k,l \in \{1,2,\cdots,K\},k\neq l,
\end{equation}
which means that the channels of different users are orthogonal to each other. Assuming that $\|\boldsymbol{h}_1\|_2 > \|\boldsymbol{h}_2\|_2 >\cdots> \|\boldsymbol{h}_K\|_2$, we can derive
\begin{align}\label{Heigen}
	\boldsymbol{H}^{\rm H}\boldsymbol{H} = \sum_{k=1}^{K} \|\boldsymbol{h}_k\|_2^2 \frac{\boldsymbol{h}_k\boldsymbol{h}_k^{\rm H}}{\|\boldsymbol{h}_k\|_2^2}  \overset{(a)}{=} \sum_{k=1}^{K} \overline{\lambda}_k [\boldsymbol{V}]_{:,1:K}[\boldsymbol{V}]_{:,1:K}^{\rm H}.
\end{align}
Note that (a) in \eqref{Heigen} holds due to the eigenvalue decomposition of $\boldsymbol{H}^{\rm H}\boldsymbol{H}$ and $\overline{\lambda}_k$ is the $k$th largest eigenvalue for $k=1,2,\cdots,K$. From \eqref{Heigen} we can find that $\overline{\lambda}_k = \|\boldsymbol{h}_k\|_2^2$ and 
\begin{equation}\label{hkexpression}
	\boldsymbol{h}_k = \|\boldsymbol{h}_k\|_2 [\boldsymbol{V}]_{:,k}.
\end{equation}
Therefore, by substituting both \eqref{FRequiv} and \eqref{hkexpression} into \eqref{Intferenosie}, we can transform \eqref{Intferenosie} as
\begin{equation}\label{fenmu}
	\boldsymbol{\Upsilon} = \boldsymbol{R}_{\overline{\eta}} + {\rm Diag}(\widetilde{\boldsymbol{h}}_1\boldsymbol{R}_{\rm q}\widetilde{\boldsymbol{h}}_1^{\rm H}, \widetilde{\boldsymbol{h}}_2\boldsymbol{R}_{\rm q}\widetilde{\boldsymbol{h}}_2^{\rm H}, \cdots,\widetilde{\boldsymbol{h}}_K\boldsymbol{F}_{\rm R}\boldsymbol{R}_{\rm q}\widetilde{\boldsymbol{h}}_K^{\rm H}).
\end{equation}
Moreover, with the designed $\boldsymbol{F}_{\rm B}$ in \eqref{FB_Heur}, we can derive that
\begin{equation}\label{interf}
	\widetilde{\boldsymbol{h}}_k\boldsymbol{f}_{{\rm B,}l}  = 0,\forall k\neq l, k,l \in \{1,2,\cdots,K\}.
\end{equation}
Finally, by substituting \eqref{fenmu} and \eqref{interf} into \eqref{MI}, we can derive
\begin{equation}\label{MItransf}
	\mathcal{I}(\boldsymbol{s};\boldsymbol{y}) = \sum_{k=1}^{K}\log\Big(1+\frac{(1-\beta)^2|\boldsymbol{h}_k^{\rm H}\boldsymbol{F}_{\rm R}\boldsymbol{f}_{{\rm B},k}|^2}{\boldsymbol{h}_k^{\rm H}\boldsymbol{F}_{\rm R}\boldsymbol{R}_{\rm q}\boldsymbol{F}_{\rm R}^{\rm H}\boldsymbol{h}_k+\sigma_k^2}\Big) = R_{\rm sum}.
\end{equation}

Proposition 2 is therefore proved. $\hfill\blacksquare$

\emph{Remark:} According to \eqref{AoDneq}, we can see that mmWave channels and massive antenna arrays inherently provide the covertness. Furthermore, with hybrid beamformer designed by the VSH scheme, the SCR can approach the mutual information asymptotically. Additionally, we can optimize the remaining parameters, i.e., $p_k$ for $k=1,2,\cdots,K$, to further improve the SCR. The detailed procedures are presented in the following subsection.

\subsection{Power Allocation Design}\label{S4subB}
After the derivation of analog and digital beamforming design as \eqref{FR_Heur} and \eqref{FB_Heur}, respectively, there still exists power allocation factors $p_k$ to be optimized. Given \eqref{FR_Heur} and \eqref{FB_Heur}, we first consider the power constraint, where the left hand in \eqref{powercstr} can be transformed as
\begin{align}\label{power_heur}
	&{\rm Tr}\Big(\boldsymbol{F}_{\rm R}\big((1-\beta)^2\sum_{l=1}^{K}\boldsymbol{f}_{{\rm B},l}\boldsymbol{f}_{{\rm B},l}^{\rm H}+\boldsymbol{R}_{\rm q}\big)\boldsymbol{F}_{\rm R}^{\rm H}\Big) \nonumber \\
	&\overset{(a)}{=}{\rm Tr}\big((1-\beta)^2\sum_{l=1}^{K}\boldsymbol{f}_{{\rm B},l}\boldsymbol{f}_{{\rm B},l}^{\rm H}+\beta(1-\beta)\sum_{l=1}^{K}{\rm diag}(\boldsymbol{f}_{{\rm B},l}\boldsymbol{f}_{{\rm B},l}^{\rm H})\big) \nonumber\\
	&={\rm Tr}\big((1-\beta)\sum_{l=1}^K p_l {\rm diag}(\widetilde{\boldsymbol{V}}_l^{(0)}\widetilde{\boldsymbol{V}}_l^{(0){\rm H}})\big) \nonumber\\
	&= (1-\beta)\sum_{l=1}^K p_l.
\end{align}
Note that $(a)$ in \eqref{power_heur} holds due to $\boldsymbol{F}_{\rm R}^{\rm H}\boldsymbol{F}_{\rm R} = \boldsymbol{I}_{K}$ via \eqref{FR_Heur}. Then, by substituting \eqref{FR_Heur}, \eqref{FB_Heur} and \eqref{power_heur} into \eqref{goalformulation_transformed}, we can transform \eqref{goalformulation_transformed} into a power allocation optimization problem, which can be expressed as
\begin{subequations}\label{powerallocationproblem}
	\begin{align}
		\underset{\overset{\{p_k\}}{k=1,2,\cdots,K}}{\max}\ &\!\sum_{k=1}^{K}\log\!\Big(\!1+\!\frac{(1-\beta)^2|\widetilde{\boldsymbol{h}}_k\widetilde{\boldsymbol{V}}_k^{(0)}|^2 p_k}{\sum_{l=1}^K\!\widetilde{\boldsymbol{V}}_l^{(0){\rm H}}{\rm diag}(\widetilde{\boldsymbol{h}}_k^{\rm H}\widetilde{\boldsymbol{h}}_k)\widetilde{\boldsymbol{V}}_l^{(0)} p_l\!+\!\sigma_k^2}\!\Big) \\
		\mathrm{s.t.\ }~~~~&(1-\beta)\sum_{l=1}^K p_l \leq P_{\rm max}.
	\end{align}
\end{subequations}

Unfortunately, we cannot straightforwardly solve \eqref{powerallocationproblem} by the popular water-filling method, due to the term $\sum_{l=1}^K\!\widetilde{\boldsymbol{V}}_l^{(0){\rm H}}{\rm diag}(\widetilde{\boldsymbol{h}}_k^{\rm H}\widetilde{\boldsymbol{h}}_k)\widetilde{\boldsymbol{V}}_l^{(0)} p_l\!$ caused by the quantization noise. To proceed, we turn to optimizing $\boldsymbol{\zeta} \triangleq [\zeta_1,\zeta_2,\cdots,\zeta_K] \in \mathbb{R}^K$ instead. Specifically, we denote $c_{l,k} \triangleq \widetilde{\boldsymbol{V}}_l^{(0){\rm H}}{\rm diag}(\widetilde{\boldsymbol{h}}_k^{\rm H}\widetilde{\boldsymbol{h}}_k)\widetilde{\boldsymbol{V}}_l^{(0)}$ for $\forall l,k\in \{1,2,\cdots,K\}$ and \eqref{powerallocationproblem} can be transformed as
\begin{subequations}\label{lograyleigh}
	\begin{align}
		\underset{\boldsymbol{\zeta}}{\max}\ &\sum_{k=1}^{K}\log\Big(1+\frac{\boldsymbol{\zeta}^{\rm T}\boldsymbol{\Pi}_k\boldsymbol{\zeta}}{\boldsymbol{\zeta}^{\rm T}\boldsymbol{\Theta}_k\boldsymbol{\zeta}+\sigma_k^2}\Big), \label{lograyleigh:sub1}\\
		\mathrm{s.t.\ }~&\boldsymbol{\zeta}^{\rm T}\boldsymbol{\zeta} \leq \frac{P_{\rm max}}{1-\beta},\label{lograyleigh:sub2}
	\end{align}
\end{subequations}
where 
\begin{align}
	&\boldsymbol{\Theta}_k = \beta(1-\beta){\rm Diag} (c_{1,k},c_{2,k},\cdots,c_{K,k}),  \\
	&\boldsymbol{\Pi}_k = (1-\beta)^2{\rm Diag}(\overbrace{0,\cdots,0}^{k-1}, |\boldsymbol{h}_k^{\rm H}\boldsymbol{F}_{\rm R}\widetilde{\boldsymbol{V}}_k^{(0)}|^2, \overbrace{0,\cdots,0}^{K-k}).                                                                                                        
\end{align}
Similar to \eqref{goalformulation_transformed}, we introduce auxiliary variables $\boldsymbol{\varrho}\triangleq [\varrho_1,\varrho_2,\cdots,\varrho_K]^{\rm T} \in \mathbb{R}^K$ and $\boldsymbol{\vartheta}\triangleq [\boldsymbol{\vartheta}_1,\boldsymbol{\vartheta}_2,\cdots,\boldsymbol{\vartheta}_K]^{\rm T}\in\mathbb{R}^{K\times K}$ so that the equivalent form of objective function \eqref{lograyleigh:sub1} can be given by
\begin{align}\label{fp}
	f_{\rm p}(\boldsymbol{\varrho},\boldsymbol{\vartheta},\boldsymbol{\zeta}) &= \sum_{k=1}^K \Big(\log(1+\varrho_k)-\varrho_k + 2\sqrt{1+\varrho_k} \boldsymbol{\vartheta}_k^{\rm T}\boldsymbol{\Pi}_k^{\frac{1}{2}}\boldsymbol{\zeta} \nonumber\\
	& ~~~- \boldsymbol{\vartheta}_k^{\rm T}\big(\boldsymbol{\zeta}^{\rm T}(\boldsymbol{\Pi}_k+\boldsymbol{\Theta}_k)\boldsymbol{\zeta}+\sigma_k^2\boldsymbol{I}_K\big)\boldsymbol{\vartheta}_k\Big).
\end{align}

Therefore, we can optimize the variables $\boldsymbol{\varrho}$, $\boldsymbol{\vartheta}$, and $\boldsymbol{\zeta}$ iteratively. Then, the optimization subproblems are given as follows.

\emph{1) Optimization for $\boldsymbol{\varrho}$:} Given $\boldsymbol{\vartheta}$ and $\boldsymbol{\zeta}$, we can derive the optimal $\widehat{\varrho}_k$ by letting $\partial f_p/\partial \varrho_k = 0$ for $k = 1,2,\cdots,K$ and it can be expressed as
\begin{equation}\label{varrho}
	\widehat{\varrho}_k = \frac{\boldsymbol{\zeta}^{\rm T}\boldsymbol{\Pi}_k\boldsymbol{\zeta}}{\boldsymbol{\zeta}^{\rm T}\boldsymbol{\Theta}_k\boldsymbol{\zeta}+\sigma_k^2}, k = 1,2,\cdots,K.
\end{equation}

\emph{2) Optimization for $\boldsymbol{\vartheta}$:} Given $\boldsymbol{\varrho}$ and $\boldsymbol{\zeta}$, we can obtain the optimal $\widehat{\vartheta}_k$ by letting $\partial f_p/\partial \vartheta_k = \boldsymbol{0}$ for $k = 1,2,\cdots,K$ and it can be given by
\begin{equation}\label{vartheta}
	\widehat{\vartheta}_k = \sqrt{1+\varrho_k}(\boldsymbol{\zeta}^{\rm T}(\boldsymbol{\Pi}_k+\boldsymbol{\Theta}_k)\boldsymbol{\zeta}+\sigma_k^2\boldsymbol{I}_K)^{-1}\boldsymbol{\Pi}_k^{\frac{1}{2}}\boldsymbol{\zeta}.
\end{equation}

\emph{3) Optimization for $\boldsymbol{\zeta}$:} Given $\boldsymbol{\varrho}$ and $\boldsymbol{\vartheta}$, the optimization problem for $\boldsymbol{\zeta}$ can be expressed as
\begin{align}\label{Zetaproblem}
	\underset{\boldsymbol{\zeta}}{\max}\ &\sum_{k=1}^K (2\sqrt{1+\varrho_k}\boldsymbol{\vartheta_k}^{\rm T}\boldsymbol{\Pi}^{\frac{1}{2}})\boldsymbol{\zeta} - \boldsymbol{\zeta}^{\rm T}\sum_{k=1}^K\big(\boldsymbol{\vartheta}_k^{\rm T}\boldsymbol{\vartheta}_k(\boldsymbol{\Pi}_k+\boldsymbol{\Theta}_k)\big)\boldsymbol{\zeta} \nonumber\\
	\mathrm{s.t.\ }~&\eqref{lograyleigh:sub2}.
\end{align}
The optimal solution of $\boldsymbol{\zeta}$ to \eqref{Zetaproblem} can be obtained by Lagrangian multiplier method. We introduce a dual variable $\mu,\mu\geq 0$ for the power constraint \eqref{lograyleigh:sub2} and derive the Lagrangian function as
\begin{align}
	\mathcal{L}(\boldsymbol{\zeta},\mu) &= \mu(\frac{P_{\rm max}}{1-\beta}-\boldsymbol{\zeta}^{\rm T}\boldsymbol{\zeta}) + \sum_{k=1}^K (2\sqrt{1+\varrho_k}\boldsymbol{\vartheta_k}^{\rm T}\boldsymbol{\Pi}^{\frac{1}{2}})\boldsymbol{\zeta} \nonumber\\
	&~~~- \boldsymbol{\zeta}^{\rm T}\sum_{k=1}^K\big(\boldsymbol{\vartheta}_k^{\rm T}\boldsymbol{\vartheta}_k(\boldsymbol{\Pi}_k+\boldsymbol{\Theta}_k)\big)\boldsymbol{\zeta}.
\end{align}
Then the optimal solution of $\boldsymbol{\zeta}$ can be determined by letting $\partial \mathcal{L}/\partial \boldsymbol{\zeta} = \boldsymbol{0}$ and it can be expressed as
\begin{equation}\label{zetaopt}
	\widehat{\boldsymbol{\zeta}} = \Big(\widehat{\mu}\boldsymbol{I}_K+\sum_{k=1}^K\big(\boldsymbol{\vartheta}_k^{\rm T}\boldsymbol{\vartheta}_k(\boldsymbol{\Pi}_k+\boldsymbol{\Theta}_k)\big)\Big)^{-1}\sum_{k=1}^K (\sqrt{1+\varrho_k}\boldsymbol{\Pi}^{\frac{1}{2}}\boldsymbol{\vartheta_k}),
\end{equation}
where $\widehat{\mu}$ can be obtained by the bisection method for a minimum $\mu,\mu\geq 0$ to satisfy $\widehat{\boldsymbol{\zeta}}(\mu)^{\rm T}\widehat{\boldsymbol{\zeta}}(\mu) \leq \frac{P_{\rm max}}{1-\beta}$. Therefore, the optimal digital beamformer can be expressed as
\begin{equation}
	\widehat{\boldsymbol{F}}_{\rm B} = [\widehat{\zeta}_1\widetilde{\boldsymbol{V}}_1^{(0)},\widehat{\zeta}_2\widetilde{\boldsymbol{V}}_2^{(0)},\cdots,\widehat{\zeta}_K\widetilde{\boldsymbol{V}}_K^{(0)}].
\end{equation}

Considering the constant-modulus constraints \eqref{goalformulation:sub2}, we can express the optimal solution approaching \eqref{FR_Heur} as~\cite{Wu2018Hysbd}
\begin{equation}\label{FRCM}
	\widehat{\boldsymbol{F}}_{\rm R} = {\rm exp}\big(j\angle\widetilde{\boldsymbol{F}}_{\rm R}\big).
\end{equation}
The approximation error of $\widehat{\boldsymbol{F}}_{\rm R}$ may cause violation of constraints. Therefore, we finally implement a scale for $\widehat{\boldsymbol{F}}_{\rm B}$ to satisfy \eqref{powercstr} and \eqref{covertcstr} and it can be expressed as
\begin{equation}\label{FBNM}
	\overline{\boldsymbol{F}}_{\rm B} = {\rm min}(1,\aleph_1,\aleph_2)\widehat{\boldsymbol{F}}_{\rm B},
\end{equation}
where $\aleph_1 = P_{\rm max}^{\frac{1}{2}}  {\rm Tr}\Big(\widehat{\boldsymbol{F}}_{\rm R}\big((1-\beta)^2\widehat{\boldsymbol{F}}_{\rm B}\widehat{\boldsymbol{F}}_{\rm B}^{\rm H}+\boldsymbol{R}_{\rm q}\big)\widehat{\boldsymbol{F}}_{\rm R}^{\rm H}\Big)^{-\frac{1}{2}}$, $\aleph_2 = \sqrt{2} \epsilon^{\frac{1}{2}} \sigma_{\rm w} T^{\frac{1}{4}}{\rm Tr}\Big(\widehat{\boldsymbol{F}}_{\rm R}\big((1-\beta)^2\widehat{\boldsymbol{F}}_{\rm B}\widehat{\boldsymbol{F}}_{\rm B}^{\rm H}+\boldsymbol{R}_{\rm q}\big)\widehat{\boldsymbol{F}}_{\rm R}^{\rm H}\boldsymbol{\Omega}_{\rm w}\Big)^{-\frac{1}{2}}$ denote the normalization factors satisfying \eqref{powercstr} and \eqref{covertcstr}, respectively. The overall VSH scheme for hybrid beamforming design is summarized in \textbf{Algorithm}~\ref{alg3}.

\begin{algorithm}[!t]
	\caption{VSH Scheme for Hybrid Beamforming Design}
	\label{alg3}
	\begin{algorithmic}[1]
		\STATE \textbf{Input}: $\boldsymbol{h}_1,\cdots,\boldsymbol{h}_K$, $\boldsymbol{\Omega}_{\rm w}$, $P_{\rm max}$, $\epsilon$.
		\STATE Initialize $\zeta_1, \zeta_2, \cdots, \zeta_K \leftarrow \sqrt{P_{\rm max}}/\sqrt{(1-\beta)K}$.
		\STATE Update analog beamformer $\boldsymbol{F}_{\rm R}$ via \eqref{FR_Heur}.
		\STATE Update digital beamformer $\boldsymbol{F}_{\rm B}$ via \eqref{FB_Heur}.
		\REPEAT
		\STATE Update $\varrho_k$ for $k = 1,2,\cdots,K$ via \eqref{varrho}.
		\STATE Update $\boldsymbol{\vartheta}_k$ for $k = 1,2,\cdots,K$ via \eqref{vartheta}.
		\STATE Update $\boldsymbol{\zeta}$ via \eqref{zetaopt}.
		\UNTIL{$f_{\rm p}$ in \eqref{fp} is converged.}
		\STATE Update $\boldsymbol{F}_{\rm R}$ via \eqref{FRCM}.
		\STATE Normalize $\boldsymbol{F}_{\rm B}$ via \eqref{FBNM}.
		\STATE \textbf{Output}: $\boldsymbol{F}_{\rm R}$, $\boldsymbol{F}_{\rm B}$.
	\end{algorithmic}
\end{algorithm}

\subsection{Complexity Analysis}\label{S4subC}
In this subsection, we analyze the computational complexity of the proposed VSH scheme in \textbf{Algorithm}~\ref{alg3}. It is seen that each step is implemented with a closed-form solution. The dominant computational cost is in steps 3, 4, 6, 7 and 8. Specifically, the SVD takes up the most computational cost in steps 3 and 4, i.e., $\mathcal{O}(N^3)$ and $\mathcal{O}(K^3)$, respectively. Step 7 takes up the most computational cost among steps 6, 7, 8 where all the computations for $K$ auxiliary variables involve matrix inversion and product and the complexity can be approximated by $\mathcal{O}(K^4)$. Therefore, supposing the iteration number for convergence in \textbf{Algorithm}~\ref{alg3} is $N_{\rm max}^{(3)}$, we can approximately obtain the overall computational complexity of the proposed VSH scheme as $\mathcal{O}(N^3 + N_{\rm max}^{(3)}K^4)$. Since the VSH scheme only has one iteration loop and needs no extra computational complexity for the initialization, it has lower computational complexity than the AO scheme.

\section{Simulation Results}\label{section5}
In this section, we evaluate the performance of the proposed schemes. We suppose that Alice is equipped with $N = 64$ antennas and $N_{\rm RF} = 4$ RF chains serving $K = 4$ legitimate users. The channels between Alice and each legitimate user (or Willie) are all established with $D=3$ channel paths with a LoS path and two NLoS paths. Specifically, we assume that the channel gain of the LoS path obeys $\alpha^{(0)}\sim \mathcal{CN}(0,1)$, the other two channel gains of NLoS paths obey $\alpha^{(1)}\sim \mathcal{CN}(0,0.01)$ and $\alpha^{(2)}\sim \mathcal{CN}(0,0.001)$~\cite{Chen2024XMIMO}. All the channel AoDs randomly distribute within $[-1,1]$. The number of time slots for Willie's detection is set as $T=100$. The noise powers are set as $\sigma_1^2=\sigma_2^2=\cdots=\sigma_K^2=\sigma_{\rm w}^2=10$ dBm. In the following we set $N = 64$, $P_{\rm max} = 0$ dBw and $\epsilon = 0.1$ as default values unless specified.

\begin{figure}[!t]
	\begin{center}
		\includegraphics[width=80mm]{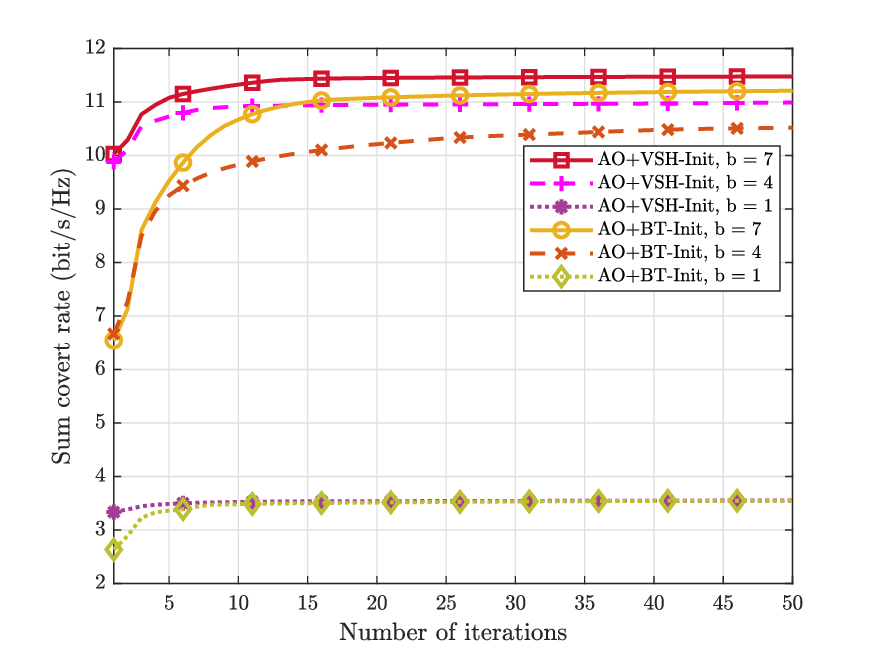}
	\end{center}
	\caption{Comparison of the SCR for different methods with varying iterations.}\label{fig:diffiterations}
	\vspace{-0.2cm}
\end{figure}

Fig.~\ref{fig:diffiterations} illustrates the convergence performance of the proposed AO scheme by plotting the SCR performance versus the iteration numbers for three cases including $b = $ 1, 4 and 7. For performance comparisons, we consider two initialization methods: 1) The beam training scheme for initialization (BT-Init) that we design the $k$th column of analog beamformer $\boldsymbol{F}_{\rm R}$ using the normalized array response in \eqref{arraryresponse} precisely towards the LoS path of the channel between Alice and the $k$th user for $k = 1,2,\cdots,K$ and employ zero-forcing method for the digital beamforming design to eliminate multiuser interference~\cite{Qi2020HBT}. 2) The VSH scheme for initialization (VSH-Init) shown in \textbf{Algorithm}~\ref{alg3}. It can be observed from Fig.~\ref{fig:diffiterations} that all the curves can achieve the convergence. Specifically, the SCR converges in approximately 15 iterations with the VSH-Init, which has a faster convergence speed than the BT-Init. Furthermore, the AO scheme with the VSH-Init can achieve better converged SCR performance than that with the BT-Init. Besides, for the three schemes including the VSH scheme, the AO scheme with the VSH-Init and the AO scheme with the BT-Init, we compare their computational complexity by measuring their running time under the same hardware and software platform. Here we set the stop condition that the relative change of the SCR within one iteration is lower than 0.001. We take $b = 7$ as an example. To achieve the convergence, the VSH scheme, the AO scheme with the VSH-Init and the AO scheme with the BT-Init need 0.0049 s, 2.0957 s, and 2.6765 s, respectively, implying that the VSH scheme has much lower computational complexity than the AO scheme. These results validate that the VSH scheme can be used to obtain an initialization for the AO scheme.

\begin{figure}[!t]
	\begin{center}
		\includegraphics[width=80mm]{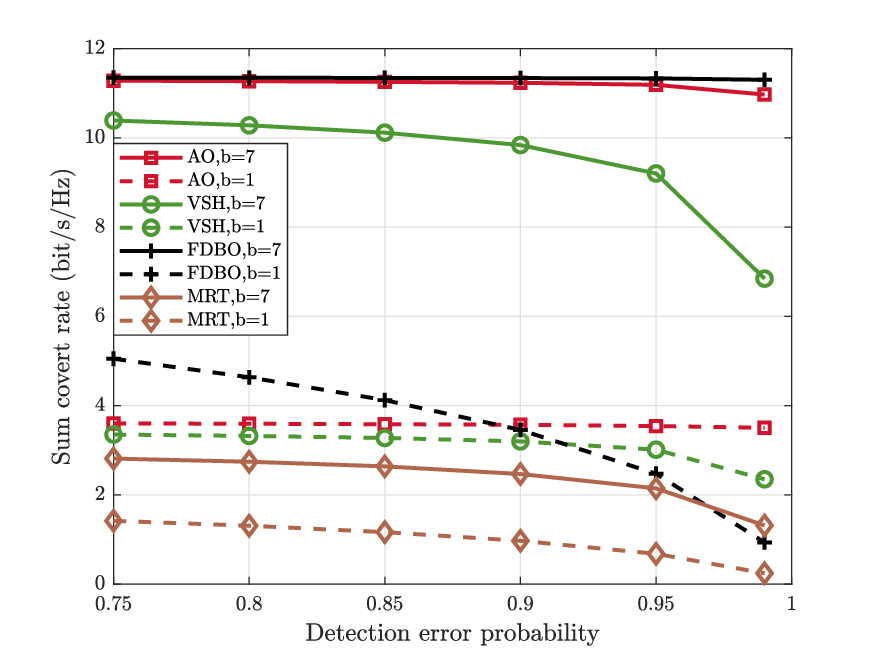}
	\end{center}
	\caption{Comparison of the SCR for different methods with varying detection error probabilities.}\label{fig:diffepsilon}
	\vspace{-0.2cm}
\end{figure}

Fig.~\ref{fig:diffepsilon} illustrates the SCR versus different predetermined detection error probabilities, and we also compare the performance of our proposed AO and VSH schemes for two cases including $b=$ 1 and 7. For comparisons, we consider two baseline schemes: 1) Fully-digital beamforming optimization (FDBO) scheme that we design a fully-digital beamformer $\boldsymbol{F}\in \mathbb{C}^{N\times K}$ with a revised AO scheme in \textbf{Algorithm}~\ref{alg2} by removing the analog beamformer design and revising the dimension of the digital beamformer. 2) Maximum ratio transmitting (MRT) scheme that we design a fully-digital beamformer $\boldsymbol{F}\in \mathbb{C}^{N\times K}$ commonly used for the covertness communication analysis~\cite{Lin2022covertMRT,Lv2022covertMRT}, where $\boldsymbol{F} = \sqrt{p}\boldsymbol{H}^{\rm H}$ and $p$ is normalized to satisfy both power and covertness constraint. It can be observed that both our proposed schemes outperform the MRT scheme. Additionally, the AO scheme always outperforms the VSH scheme due to the approximation errors caused by assuming that $N$ approaches infinity in the VSH scheme. Furthermore, when the quantization noise is negligible at $b = 7$, the AO scheme can achieve performance comparable to that of the FDBO scheme. Moreover, the AO scheme demonstrates superior stability against variations in covertness constraints compared to the FDBO scheme when $b = 1$. The reason is that in the hybrid beamforming architecture, the analog beamforming reduces the leakage of quantization noise in the Alice-Willie link, making it easier to satisfy the covertness constraint. Therefore, these results validate the effectiveness of hybrid beamforming design with finite-resolution DACs in real-world covert scenarios.

\begin{figure}[!t]
	\begin{center}
		\includegraphics[width=80mm]{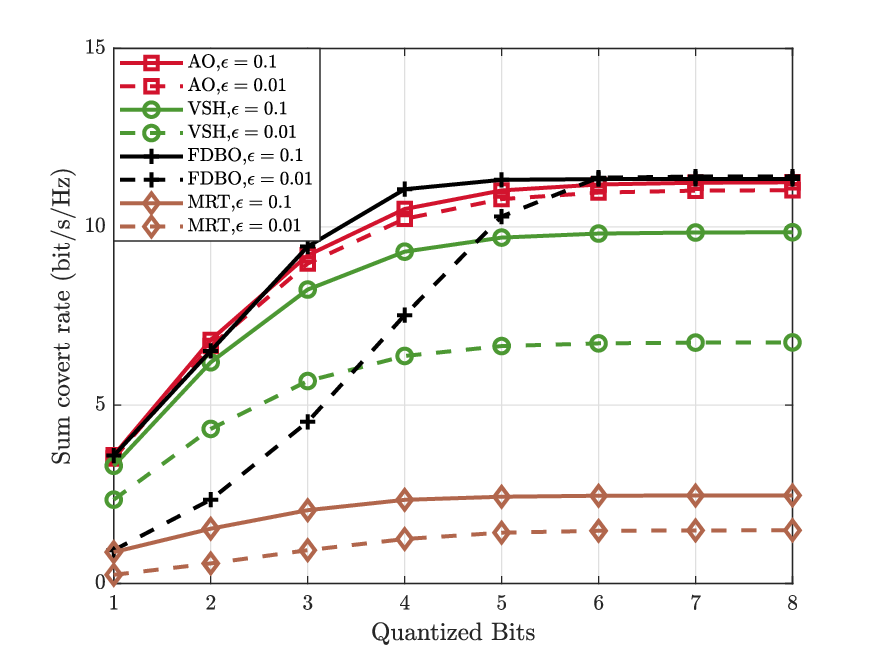}
	\end{center}
	\caption{Comparison of the SCR for different methods with varying quantized bits.}\label{fig:diffbit}
	\vspace{-0.2cm}
\end{figure}

In Fig.~\ref{fig:diffbit}, we evaluate the SCR performance as a function of $b$ to illustrate the impact of the quantized bits of DACs. It can be observed that as $b$ increases, the SCR performance first improves when $b \leq 5$ and then tends to be flat when $b > 5$. This phenomenon occurs because that the quantization noise decreases rapidly with increasing $b$. Additionally, the result shows that our proposed AO scheme exhibits lower sensitivity to the covertness compared to the FDBO scheme. When $\epsilon = 0.01$ and $b \leq 5$, our proposed AO scheme outperforms the FDBO scheme. The reason is that the analog beamformer in the hybrid beamforming architecture reduces the leakage of quantization noise in the Alice-Willie link so that the transmit power can be utilized as fully as possible without exceeding $P_{\rm max}$ to achieve better SCR performance. Besides, the AO scheme also achieves nearly the same SCR performance as the FDBO scheme when $b >5$. 

\begin{figure}[!t]
	\begin{center}
		\includegraphics[width=80mm]{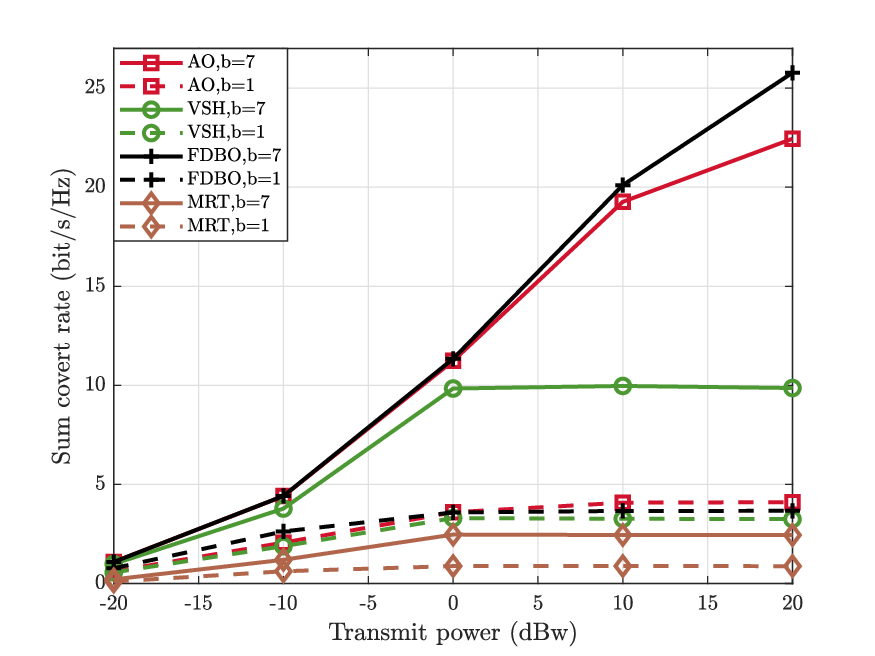}
	\end{center}
	\caption{Comparison of the SCR for different methods with varying transmit power.}\label{fig:diffPcmax}
	\vspace{-0.2cm}
\end{figure}

Fig.~\ref{fig:diffPcmax} illustrates the SCR performance with respect to $P_{\rm max}$. It can be observed that the SCR initially increases and then tends to be flat with the growth of $P_{\rm max}$, especially for the case of low-resolution DAC, e.g., $b=1$. This result can be explained as follows. During the ascending stage of the curve, $P_{\rm max}$ predominantly restricts the SCR performance. In the subsequent stage when the curves tend to be flat, the covertness primarily restricts the SCR performance.

\begin{figure}[!t]
	\begin{center}
		\includegraphics[width=80mm]{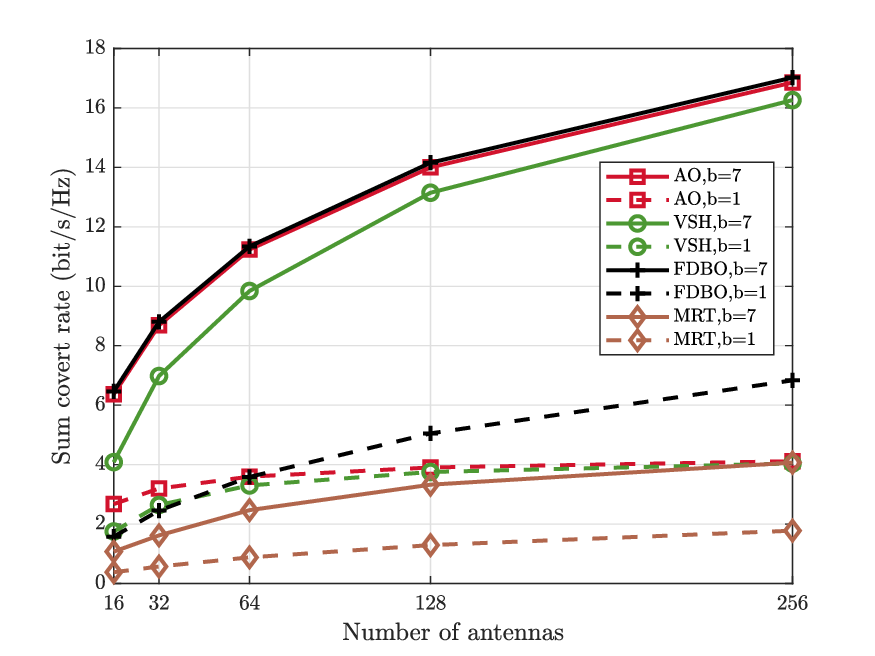}
	\end{center}
	\caption{Comparison of the SCR for different methods with varying numbers of antennas.}\label{fig:diffantennas}
	\vspace{-0.2cm}
\end{figure}

In Fig.~\ref{fig:diffantennas}, we compare the SCR performance in terms of the number of transmit antennas. As $N$ increases, the SCR always increases due to the enhanced beamforming gain achieved by a larger antenna array. Furthermore, in the case of low-resolution DAC, e.g., $b=1$, both the AO and VSH schemes outperform the MRT scheme. Additionally, the SCR obtained by the FDBO scheme increases fastest along with increasing $N$ among all the schemes, due to the enhanced design flexibility of the digital beamformers than the analog beamformers. However, this advantage is achieved with the price of consuming more power. In the case of high-resolution DAC, e.g., $b=7$, our proposed schemes achieve nearly the same SCR performance as that of the FDBO scheme. Moreover, Fig.~\ref{fig:diffantennas} demonstrates that as $N$ increases, the gaps in the SCR performance between the AO and VSH schemes decrease, validating the effectiveness of the VSH scheme in the massive MIMO systems.
 
\begin{figure}[!t]
	\begin{center}
		\includegraphics[width=80mm]{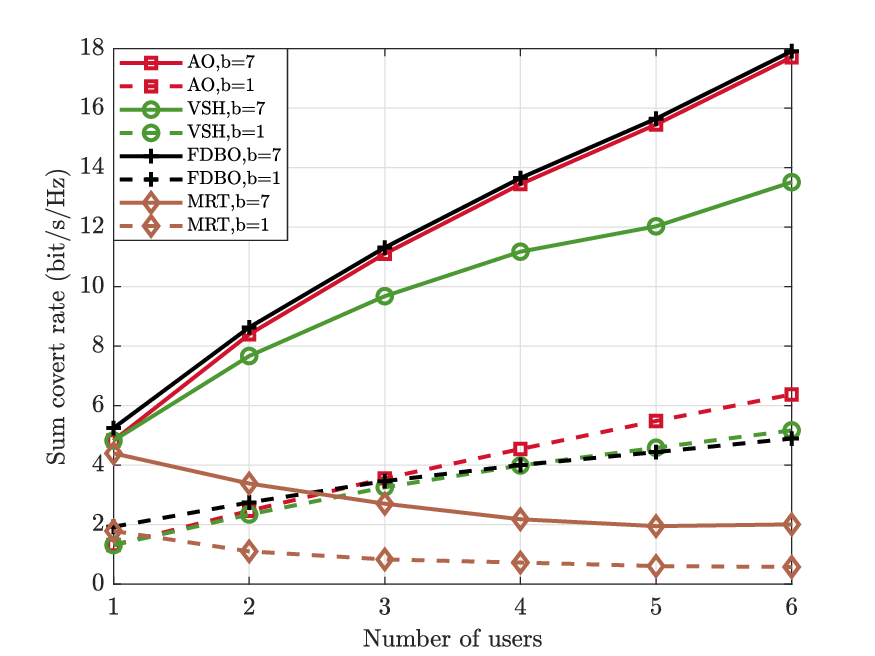}
	\end{center}
	\caption{Comparison of the SCR for different methods with varying numbers of users.}\label{fig:diffUsers}
	\vspace{-0.2cm}
\end{figure}

Fig.~\ref{fig:diffUsers} illustrates the SCR performance versus different numbers of legitimate users. As $K$ increases, all the FDBO, the AO and the VSH schemes demonstrate an improvement in SCR performance, whereas the MRT scheme exhibits a decline. The reason is that the MRT scheme does not consider the covertness, multi-user interference, and quantization noise suppression, different from the other three schemes. Furthermore, the AO and the VSH schemes outperform the FDBO scheme when $K > 3$ in the case of low-resolution DACs, e.g., $b=1$, due to smaller quantization noise in the hybrid beamforming architecture. In contrast, for the case of high-resolution DAC, e.g, $b=7$, the gaps in the SCR performance between the AO and VSH schemes increase with the growth of $K$, primarily due to the more strict covertness requirements for larger $K$. However, the VSH scheme is still valuable for its low complexity and moderate performance.

Finally, we evaluate the performance of energy efficiency to investigate the tradeoff between the SCR and power consumption. We define the energy efficiency as
\begin{equation}
	\chi \triangleq \frac{R_{\rm sum}}{P_{\rm tot}},
\end{equation}
where $P_{\rm tot}$ is the total power consumption of the transmitter. We denote $P_{\rm c}$, $P_{\rm LNA}$, $P_{\rm PS}$, $P_{\rm RF}$, $P_{\rm DAC}$ and $P_{\rm BB}$ as transmit power, power consumption of the low noise amplifier, phase shifter, RF chain, DAC and baseband processor, respectively. Then, the total power consumption with the fully-digital transmitter can be approximated by
\begin{equation}
	P_{\rm tot}^{\rm FD} \triangleq P_{\rm c} + N(P_{\rm LNA}+P_{\rm RF}+2 P_{\rm DAC})+P_{\rm BB}.
\end{equation}
Similarly, the total power consumption with the fully-connected hybrid beamforming architecture can be approximated by
\begin{equation}
	P_{\rm tot}^{\rm HBF} \triangleq P_{\rm c} + N P_{\rm LNA} + N_{\rm RF}(N P_{\rm PS}+P_{\rm RF}+2 P_{\rm DAC}) +P_{\rm BB}.
\end{equation}
Note that $P_{\rm c}$ can be calculated with the left hand of \eqref{powercstr}. Then, according to~\cite{Mo2017ADC}, we assume that $P_{\rm LNA} = 20$ mW, $P_{\rm PS} = 10$ mW, $P_{\rm RF} = 40$ mW and $P_{\rm BB} = 200$ mW in our simulations. The power consumed by DAC is modeled as 
\begin{equation}
	P_{\rm DAC} = 2^b f_{\rm s} FOM_{\rm W},
\end{equation}
where $f_s$ and $FOM_{\rm W}$ are denoted as sampling rate and the DAC's power efficiency with resolution and sampling rate, respectively. Typically we set $f_s=1$ GHz and $FOM_{\rm W} = 500$ fJ/conversion-step~\cite{Mo2017ADC}. 

\begin{figure}[!t]
	\begin{center}
		\includegraphics[width=80mm]{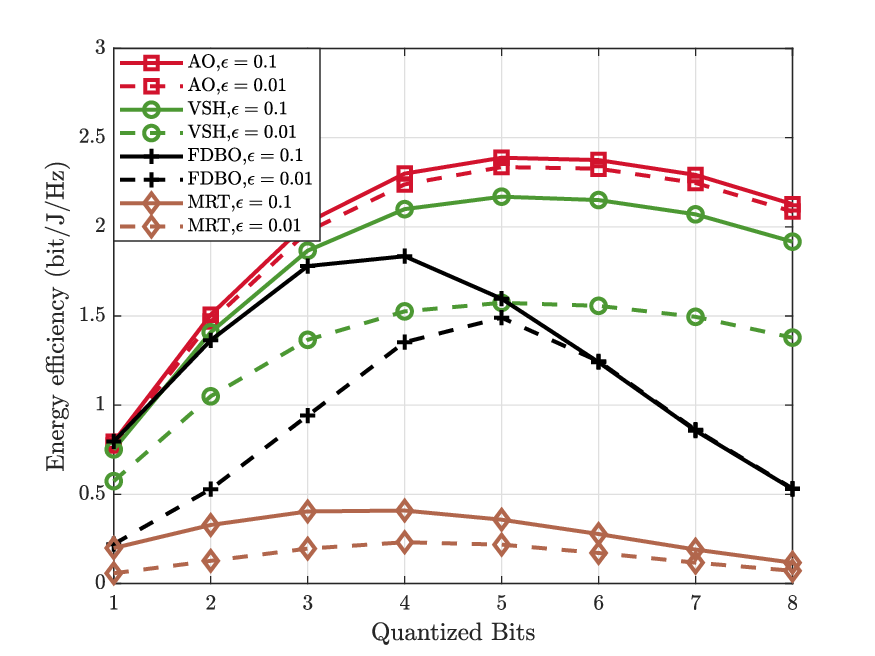}
	\end{center}
	\caption{Comparison of the energy efficiency for different methods with varying quantized bits.}\label{fig:diffbit_EE}
	\vspace{-0.2cm}
\end{figure}

The energy efficiency performance versus the quantized bits $b$ of DACs is illustrated in Fig.~\ref{fig:diffbit_EE}. It can be observed that the energy efficiency does not always increase with the growth of $b$, and an optimal $b$ for maximizing the energy efficiency performance exists. Furthermore, our proposed two schemes outperform the FDBO and MRT schemes for the energy efficiency performance. Additionally, the optimal $b$ depends on different predetermined covertness requirements. This result emphasizes the significance of the DAC resolutions for practical covert mmWave MIMO communications.

\section{conclusion}\label{section6}
In this paper, we have investigated the hybrid beamforming design for mmWave covert MIMO communication systems with finite-resolution DACs. We have proposed the AO scheme to maximize the SCR, where both analog and digital beamformers are optimized iteratively. Specifically, the analog beamforming design has been obtained using an iMM method, while the digital beamforming design has been solved via the interior-point method. To reduce the computational complexity of the AO scheme, we have proposed the VSH scheme for hybrid beamforming design, which can also obtain a initialization for the AO scheme. For future research, it would be interesting to extend our work to the wideband mmWave MIMO communications.

\bibliographystyle{IEEEtran} 
\bibliography{IEEEabrv,IEEEexample}

\begin{thebibliography}{10}
\providecommand{\url}[1]{#1}
\csname url@samestyle\endcsname
\providecommand{\newblock}{\relax}
\providecommand{\bibinfo}[2]{#2}
\providecommand{\BIBentrySTDinterwordspacing}{\spaceskip=0pt\relax}
\providecommand{\BIBentryALTinterwordstretchfactor}{4}
\providecommand{\BIBentryALTinterwordspacing}{\spaceskip=\fontdimen2\font plus
\BIBentryALTinterwordstretchfactor\fontdimen3\font minus
  \fontdimen4\font\relax}
\providecommand{\BIBforeignlanguage}[2]{{%
\expandafter\ifx\csname l@#1\endcsname\relax
\typeout{** WARNING: IEEEtran.bst: No hyphenation pattern has been}%
\typeout{** loaded for the language `#1'. Using the pattern for}%
\typeout{** the default language instead.}%
\else
\language=\csname l@#1\endcsname
\fi
#2}}
\providecommand{\BIBdecl}{\relax}
\BIBdecl

\bibitem{Ci2024Finite}
W.~Ci and C.~Qi, ``Finite-resolution {DACs} based hybrid beamforming design for
  covert communications,'' in \emph{Proc. 2024 {lEEE} Global Communications
  Conference (GLOBECOM)}, Cape Town, South Africa, Dec. 2024, pp. 1--6.

\bibitem{Hu2021DRL}
Q.~Hu, Y.~Liu, Y.~Cai, G.~Yu, and Z.~Ding, ``Joint deep reinforcement learning
  and unfolding: Beam selection and precoding for {mmWave} multiuser {MIMO}
  with lens arrays,'' \emph{{IEEE} J. Sel. Areas Commun.}, vol.~39, no.~8, pp.
  2289--2304, Aug. 2021.

\bibitem{Qi2020HBT}
C.~Qi, K.~Chen, O.~A. Dobre, and G.~Y. Li, ``Hierarchical codebook-based
  multiuser beam training for millimeter wave massive {MIMO},'' \emph{{IEEE}
  Trans. Wireless Commun.}, vol.~19, no.~12, pp. 8142--8152, Dec. 2020.

\bibitem{Bash2013Limits}
B.~A. {Bash}, D.~{Goeckel}, and D.~{Towsley}, ``Limits of reliable
  communication with low probability of detection on {AWGN} channels,''
  \emph{{IEEE} J. Sel. Areas Commun.}, vol.~31, no.~9, pp. 1921--1930, Sep.
  2013.

\bibitem{Bloch2016Covert}
M.~R. Bloch, ``Covert communication over noisy channels: A resolvability
  perspective,'' \emph{{IEEE} Trans. Inf. Theory}, vol.~62, no.~5, pp.
  2334--2354, May 2016.

\bibitem{AKS2019Embedding}
K.~S. Kumar~Arumugam and M.~R. Bloch, ``Embedding covert information in
  broadcast communications,'' \emph{{IEEE} Trans. Inf. Forensics Security},
  vol.~14, no.~10, pp. 2787--2801, Oct. 2019.

\bibitem{AKS2019Covert}
K.~S.~K. Arumugam and M.~R. Bloch, ``Covert communication over a \emph{K}-user
  multiple-access channel,'' \emph{{IEEE} Trans. Inf. Theory}, vol.~65, no.~11,
  pp. 7020--7044, Nov. 2019.

\bibitem{Goeckel2016covert}
D.~Goeckel, B.~Bash, S.~Guha, and D.~Towsley, ``Covert communications when the
  warden does not know the background noise power,'' \emph{{IEEE} Commun.
  Lett.}, vol.~20, no.~2, pp. 236--239, Feb. 2016.

\bibitem{Shahzad2017Fading}
K.~{Shahzad}, X.~{Zhou}, and S.~{Yan}, ``Covert communication in fading
  channels under channel uncertainty,'' in \emph{Proc. {IEEE} 85th Veh.
  Technol. Conf. (VTC Spring)}, Sydney, NSW, Australia, June 2017, pp. 1--5.

\bibitem{Sobers2017Uninformed}
T.~V. {Sobers}, B.~A. {Bash}, S.~{Guha}, D.~{Towsley}, and D.~{Goeckel},
  ``Covert communication in the presence of an uninformed jammer,''
  \emph{{IEEE} Trans. Wireless Commun.}, vol.~16, no.~9, pp. 6193--6206, Sep.
  2017.

\bibitem{Shahzad2018Achieving}
K.~{Shahzad}, X.~{Zhou}, S.~{Yan}, J.~{Hu}, F.~{Shu}, and J.~{Li}, ``Achieving
  covert wireless communications using a full-duplex receiver,'' \emph{{IEEE}
  Trans. Wireless Commun.}, vol.~17, no.~12, pp. 8517--8530, Dec. 2018.

\bibitem{Zheng2021Wireless}
T.-X. Zheng, Z.~Yang, C.~Wang, Z.~Li, J.~Yuan, and X.~Guan, ``Wireless covert
  communications aided by distributed cooperative jamming over slow fading
  channels,'' \emph{{IEEE} Trans. Wireless Commun.}, vol.~20, no.~11, pp.
  7026--7039, Nov. 2021.

\bibitem{Ma2021Robust}
S.~Ma \emph{et~al.}, ``Robust beamforming design for covert communications,''
  \emph{{IEEE} Trans. Inf. Forensics Security}, vol.~16, pp. 3026--3038, 2021.

\bibitem{Jamali2022Covert}
M.~V. Jamali and H.~Mahdavifar, ``Covert millimeter-wave communication: Design
  strategies and performance analysis,'' \emph{{IEEE} Trans. Wireless Commun.},
  vol.~21, no.~6, pp. 3691--3704, June 2022.

\bibitem{Zhang2021Beamtraining}
J.~Zhang \emph{et~al.}, ``Joint beam training and data transmission design for
  covert millimeter-wave communication,'' \emph{{IEEE} Trans. Inf. Forensics
  Security}, vol.~16, pp. 2232--2245, 2021.

\bibitem{Xing2023covert}
Z.~Xing, C.~Qi, Y.~Cheng, Y.~Wu, D.~Lv, and P.~Li, ``Covert millimeter wave
  communications based on beam sweeping,'' \emph{{IEEE} Commun. Lett.},
  vol.~27, no.~5, pp. 1287--1291, Mar. 2023.

\bibitem{Cai2020Secure}
Y.~Cai, F.~Cui, Q.~Shi, Y.~Wu, B.~Champagne, and L.~Hanzo, ``Secure hybrid
  {A/D} beamforming for hardware-efficient large-scale multiple-antenna {SWIPT}
  systems,'' \emph{{IEEE} Trans. Commun.}, vol.~68, no.~10, pp. 6141--6156,
  Oct. 2020.

\bibitem{Chen2020twostep}
K.~Chen, C.~Qi, and G.~Y. Li, ``Two-step codeword design for millimeter wave
  massive {MIMO} systems with quantized phase shifters,'' \emph{{IEEE} Trans.
  Signal Process.}, vol.~68, pp. 170--180, Dec. 2020.

\bibitem{Wang2022covertRate}
C.~Wang, Z.~Li, and D.~W.~K. Ng, ``Covert rate optimization of millimeter wave
  full-duplex communications,'' \emph{{IEEE} Trans. Wireless Commun.}, vol.~21,
  no.~5, pp. 2844--2861, May 2022.

\bibitem{Orhan2015LowPower}
O.~Orhan, E.~Erkip, and S.~Rangan, ``Low power analog-to-digital conversion in
  millimeter wave systems: Impact of resolution and bandwidth on performance,''
  in \emph{Proc. Inf. Theory Appl. Workshop (ITA)}, San Diego, CA, USA, Feb.
  2015, pp. 191--198.

\bibitem{Abbas2017Lowresolution}
W.~B. Abbas, F.~Gomez-Cuba, and M.~Zorzi, ``Millimeter wave receiver
  efficiency: A comprehensive comparison of beamforming schemes with low
  resolution {ADCs},'' \emph{{IEEE} Trans. Wireless Commun.}, vol.~16, no.~12,
  pp. 8131--8146, Dec. 2017.

\bibitem{Xu2019Secure}
L.~Xu \emph{et~al.}, ``Secure hybrid digital and analog precoder for {mmWave}
  systems with low-resolution {DACs} and finite-quantized phase shifters,''
  \emph{IEEE Access}, vol.~7, pp. 109\,763--109\,775, 2019.

\bibitem{Wu2023secure}
D.~Wu \emph{et~al.}, ``Secure hybrid analog and digital beamforming for
  {mmWave} {XR} communications with mixed-{DAC},'' \emph{{IEEE} J. Sel. Topics
  Signal Process.}, vol.~17, no.~5, pp. 995--1006, Sep. 2023.

\bibitem{Roth2017lowbeamforming}
K.~Roth and J.~A. Nossek, ``Achievable rate and energy efficiency of hybrid and
  digital beamforming receivers with low resolution {ADC},'' \emph{{IEEE} J.
  Sel. Areas Commun.}, vol.~35, no.~9, pp. 2056--2068, Sep. 2017.

\bibitem{Fletcher2007AQNM}
A.~K. Fletcher, S.~Rangan, V.~K. Goyal, and K.~Ramchandran, ``Robust predictive
  quantization: Analysis and design via convex optimization,'' \emph{{IEEE} J.
  Sel. Topics Signal Process.}, vol.~1, no.~4, pp. 618--632, Dec. 2007.

\bibitem{Zhao2024SDMA}
X.~Zhao, W.~Deng, M.~Li, and M.-J. Zhao, ``Robust beamforming design for
  integrated sensing and covert communication systems,'' \emph{IEEE Wireless
  Commun. Lett., early access}, pp. 1--5, 2024.

\bibitem{EIT2002}
T.~M. Cover and J.~A. Thomas, \emph{Elements of Information Theory},
  2nd~ed.\hskip 1em plus 0.5em minus 0.4em\relax Hoboken, NJ, USA: Wiley, 2002.

\bibitem{Ma2022covertrand}
R.~Ma, W.~Yang, L.~Tao, X.~Lu, Z.~Xiang, and J.~Liu, ``Covert communications
  with randomly distributed wardens in the finite blocklength regime,''
  \emph{{IEEE} Trans. Veh. Technol.}, vol.~71, no.~1, pp. 533--544, Jan. 2022.

\bibitem{Shen2018quadratic}
K.~Shen and W.~Yu, ``Fractional programming for communication systems---{Part}
  ${\rm \uppercase\expandafter{\romannumeral2}}$: Uplink scheduling via
  matching,'' \emph{{IEEE} Trans. Signal Process.}, vol.~66, no.~10, pp.
  2631--2644, May 2018.

\bibitem{Qi2022Hybridswitches}
C.~Qi, Q.~Liu, X.~Yu, and G.~Y. Li, ``Hybrid precoding for mixture use of phase
  shifters and switches in {mmWave} massive {MIMO},'' \emph{{IEEE} Trans.
  Commun.}, vol.~70, no.~6, pp. 4121--4133, June 2022.

\bibitem{He2022QCQP}
X.~He and J.~Wang, ``{QCQP} with extra constant modulus constraints: Theory and
  application to {SINR} constrained {mmWave} hybrid beamforming,'' \emph{{IEEE}
  Trans. Signal Process.}, vol.~70, pp. 5237--5250, 2022.

\bibitem{Huang2022PTCISAC}
Z.~Huang, K.~Wang, A.~Liu, Y.~Cai, R.~Du, and T.~X. Han, ``Joint pilot
  optimization, target detection and channel estimation for integrated sensing
  and communication systems,'' \emph{{IEEE} Trans. Wireless Commun.}, vol.~21,
  no.~12, pp. 10\,351--10\,365, Dec. 2022.

\bibitem{Peng2021Strategies}
X.~Peng, J.~Wang, S.~Xiao, and W.~Tang, ``Strategies in covert communication
  with imperfect channel state information,'' in \emph{Proc. 2021 {IEEE} Global
  Commun. Conf. (GLOBECOM)}, Madrid, Spain, Dec. 2021, pp. 1--6.

\bibitem{Wu2018Hysbd}
X.~Wu, D.~Liu, and F.~Yin, ``Hybrid beamforming for multi-user massive {MIMO}
  systems,'' \emph{{IEEE} Trans. Commun.}, vol.~66, no.~9, pp. 3879--3891, Sep.
  2018.

\bibitem{Chen2024XMIMO}
K.~Chen, C.~Qi, C.-X. Wang, and G.~Y. Li, ``Beam training and tracking for
  extremely large-scale {MIMO} communications,'' \emph{{IEEE} Trans. Wireless
  Commun.}, vol.~23, no.~5, pp. 5048--5062, May 2024.

\bibitem{Lin2022covertMRT}
Y.~Lin, L.~Jin, K.~Huang, Z.~Zhong, and Q.~Han, ``Covert threat region analysis
  of {3-D} location-based beamforming in {Rician} channel,'' \emph{{IEEE}
  Wireless Commun. Lett.}, vol.~11, no.~6, pp. 1253--1257, June 2022.

\bibitem{Lv2022covertMRT}
L.~Lv, Z.~Li, H.~Ding, N.~Al-Dhahir, and J.~Chen, ``Achieving covert wireless
  communication with a multi-antenna relay,'' \emph{{IEEE} Trans. Inf.
  Forensics Security}, vol.~17, pp. 760--773, 2022.

\bibitem{Mo2017ADC}
J.~Mo, A.~Alkhateeb, S.~Abu-Surra, and R.~W. Heath, ``Hybrid architectures with
  few-bit {ADC} receivers: Achievable rates and energy-rate tradeoffs,''
  \emph{{IEEE} Trans. Wireless Commun.}, vol.~16, no.~4, pp. 2274--2287, Apr.
  2017.

\end{thebibliography}

\end{document}